\documentclass[longauth]{aa}
\usepackage{graphicx}
\usepackage[varg]{txfonts}
\usepackage{url}
\usepackage{threeparttable}
\usepackage{longtable}
\usepackage{siunitx}
\usepackage{balance}
\usepackage[normalem]{ulem}

\newcommand{\ktwo}{\emph{K2}}
\newcommand{\tess}{\emph{TESS}}
\newcommand{\gaia}{\emph{Gaia}}
\newcommand{\kepler}{\emph{Kepler}}
\newcommand{\corot}{\emph{CoRoT}}
\newcommand{\plato}{\emph{PLATO}}
\newcommand{\sname}{EPIC 249893012}

\newcommand\maspyr{\mathrm{mas\,yr^{-1}}}
\newcommand{\vsini}{$V \sin i_\star$}

\newcommand{\vmic}{$V_{\rm mic}$}
\newcommand{\vmac}{$V_{\rm mac}$}

\newcommand{\teff}{$T_{\rm eff}$}
\newcommand{\logg}{log\,{\it g$_\star$}}

\newcommand{\feh}{[Fe/H]}
\newcommand{\mgh}{[Mg/H]}
\newcommand{\nah}{[Na/H]}
\newcommand{\cah}{[Ca/H]}

\newcommand{\kms}{km\,s$^{-1}$}

\newcommand{\gc}{g~cm$^{-3}$}

\newcommand{\Msun}{M$_{\odot}$}
\newcommand{\Rsun}{R$_{\odot}$}

\newcommand{\cai}{Ca\,{\sc I} }

\newcommand{\mgi}{Mg\,{\sc I} }

\newcommand{\mstar}{$M_\star$}
\newcommand{\rstar}{$R_\star$}

\begin{document}
\title{Three planets transiting the evolved star EPIC\,249893012: \\ a hot 8.8-M$_\oplus$ super-Earth and two warm 14.7 and 10.2-M$_\oplus$ sub-Neptunes,\thanks{Based on observations made with the ESO-3.6m telescope at La Silla Observatory (Chile) under programs 0101.C-0829, 1102.C-0923, and 60.A-9700.},\thanks{Based on observations made with the Italian Telescopio Nazionale Galileo (TNG) operated on the island of La Palma by the Fundación Galileo Galilei of the INAF (Istituto Nazionale di Astrofisica) at the Spanish Observatorio del Roque de los Muchachos of the Instituto de Astrofisica de Canarias, under programs CAT18A\_130, CAT18B\_93, and A37TAC\_37.}
}

\titlerunning{EPIC 249893012: A transiting multi-planet system around an evolved star}

\author{
        Hidalgo,~D.\inst{1}$^,$ \inst{2},
        Pallé,~E.\inst{1}$^,$ \inst{2},
		Alonso,~R.\inst{1}$^,$ \inst{2},
		Gandolfi,~D.\inst{3},
        Fridlund,~M.\inst{4}$^,$ \inst{5},
        Nowak,~G.\inst{1}$^,$ \inst{2},
        Luque,~R.\inst{1}$^,$ \inst{2},
        Hirano,~T.\inst{6},
        Justesen,~A.\,B.\inst{11},
		Cochran,~W.\,D.\inst{12},
		Barragan,~O.\inst{3}$^,$ \inst{27},
		Spina,~L.\inst{24},
		Rodler,~F.\inst{25},
		Albrecht,~S.\inst{11},
		Anderson,~D.\inst{34}$^,$ \inst{35},
		Amado,~P.\inst{33},
		Bryant,~E.\inst{33}$^,$ \inst{34},
		Caballero,~J.\,A.\inst{30},
		Cabrera,~J.\inst{16},
		Csizmadia,~Sz.\inst{16},
		Dai,~F.\inst{13}$^,$ \inst{14},
		De Leon,~J.\inst{10},
		Deeg,~H.\,J.\inst{1}$^,$ \inst{2},
		Eigmuller,~Ph.\inst{16}$^,$ \inst{17},
		Endl,~M.\inst{12},
		Erikson,~A.\inst{16},
		Esposito,~M.\inst{18},
		Figueira,~P.\inst{36}$^,$ \inst{37},
		Georgieva,~I.\inst{4},
		Grziwa,~S.\inst{19},
		Guenther,~E.\inst{18},
		Hatzes,~A.\,P.\inst{18},
		Hjorth,~M.\inst{11},
		Hoeijmakers,~H.\,J.\inst{39}$^,$ \inst{40},
		Kabath,~P.\inst{28},
		Korth,~J.\inst{19},
		Kuzuhara,~M.\inst{9}$^,$ \inst{10},
		Lafarga,~M.\inst{23}$^,$ \inst{24},
		Lampon,~M.\inst{33},
		Le\~ao,~I. C.\inst{38},
		Livingston,~J.\inst{7},
		Mathur,~S.\inst{1}$^,$ \inst{2},
		Monta\~nes-Rodriguez,~P.\inst{1}$^,$ \inst{2},
		Morales,~J.\,C.\inst{22}$^,$ \inst{23},
		Murgas,~F.\inst{1}$^,$ \inst{2},
		Nagel,~E.\inst{21},
		Narita,~N.\inst{1}$^,$ \inst{8}$^,$ \inst{9}$^,$ \inst{10},
		Nielsen,~L.\,D.\inst{39},
		Patzold,~M.\inst{19},
		Persson,~C.\,M.\inst{4},
		Prieto-Arranz,~J.\inst{1}$^,$ \inst{2},
		Quirrenbach,~A.\inst{32},
		Rauer,~H.\inst{16}$^,$ \inst{17}$^,$ \inst{20},
		Redfield,~S.\inst{15},
		Reiners,~A.\inst{31},
		Ribas,~I.\inst{22}$^,$ \inst{23},
		Smith,~A.\,M.\,S.\inst{16},
		\v{S}ubjak,~J.\inst{28}$^,$ \inst{29},
		Van~Eylen,~V.\inst{26},
		Wilson,~P.\,A.\inst{34}$^,$ \inst{35}
        }
\authorrunning{Hidalgo, D. et al.}

\offprints{D. Hidalgo, \email{dhidalgo@iac.es}}

\institute{Instituto de Astrofisica de Canarias, Tenerife, Spain 
           \and Dpto. Astrofísica Universidad de La Laguna, Tenerife, Spain 
           \and Dipartimento di Fisica, Universit\`a degli Studi di Torino, Torino, Italy 
           \and Department of Space, Earth and Environment, Chalmers University of Technology, Onsala Space Observatory, SE-439 92 Onsala, Sweden 
           \and Leiden Observatory, University of Leiden, Leiden, The Netherlands 
           \and Department of Earth and Planetary Sciences, Tokyo Institute of Technology, 2-12-1 Ookayama, Meguro-ku, Tokyo 152-8551, Japan 
           \and Department of Astronomy, Graduate School of Science, The University of Tokyo, Hongo 7-3-1, Bunkyo-ku, Tokyo, 113-0033, Japan 
           \and JST, PRESTO, 2-21-1 Osawa, Mitaka, Tokyo 181-8588, Japan 
           \and Astrobiology Center, NINS, 2-21-1 Osawa, Mitaka, Tokyo 181-8588, Japan 
           \and National Astronomical Observatory of Japan, NINS, 2-21-1 Osawa, Mitaka, Tokyo 181-8588, Japan 
           \and Stellar Astrophysics Centre, Deparment of Physics and Astronomy, Aarhus University, Ny Munkegrade 120, DK-8000 Aarhus C, Denmark 
           \and Department of Astronomy and McDonald Observatory, University of Texas at Austin, 2515 Speedway, Stop C1400, Austin, TX 78712, USA 
           \and Department of Physics and Kavli Institute for Astrophysics and Space Research, Massachusetts Institute of Technology, Cambridge, MA, 02139, USA 
           \and Department of Astrophysical Sciences, Princeton University, 4 Ivy Lane, Princeton, NJ, 08544, USA 
           \and Astronomy Department and Van Vleck Observatory, Wesleyan University, Middletown, CT 06459, USA 
           \and Institute of Planetary Research, German Aerospace Center, Rutherfordstrasse 2, 12489 Berlin, Germany 
           \and Center for Astronomy and Astrophysics, TU Berlin, Hardenbergstr. 36, 10623 Berlin, Germany 
           \and Th\"uringer Landessternwarte Tautenburg, Sternwarte 5, D-07778 Tautenburg, Germany 
           \and Rheinisches Institut f\"ur Umweltforschung, Abteilung Planetenforschung an der Universit\"at zu K\"oln, Aachener Strasse 209, 50931 K\"oln, Germany 
           \and Institute of Geological Sciences, Freie Universit\"at Berlin, Malteserstr. 74-100, 12249 Berlin, Germany
           \and Hamburger Sternwarte, Gojenbergsweg 112, 21029 Hamburg, Germany 
           \and Institut de Ci\`encies de l'Espai (ICE, CSIC), Campus UAB, C/ de Can Magrans s/n, E-08193 Bellaterra, Spain 
           \and Institut d'Estudis Espacials de Catalunya (IEEC), C/ Gran Capit\`a 2-4, E-08034 Barcelona, Spain 
           \and Monash Centre for Astrophysics, School of Physics and Astronomy, Monash University, VIC 3800, Australia
           \and European Southern Observatory (ESO), Alonso de C\'ordova 3107, Vitacura, Casilla 19001, Santiago de Chile 
           \and  Mullard Space Science Laboratory, University College London, Holmbury St Mary, Dorking, Surrey RH5 6NT, United Kingdom 
           \and Sub-department of Astrophysics, Department of Physics, University of Oxford, Oxford OX1 3RH, UK 
           \and Astronomical Institute of the Czech Academy of Sciences, Fri\v{c}ova 298, 25165 Ond\v{r}ejov, Czech Republic 
           \and Astronomical Institute of Charles University, V Hole\v{s}ovi\v{c}k\'{a}ch 2, 180 00, Praha, Czech Republic 
           \and Centro de Astrobiolog\'ia (CSIC-INTA), ESAC, camino bajo del castillo s/n, E-28692 Villanueva de la Ca\~nada, Madrid Spain 
           \and Institut f\"ur Astrophysik, Georg-August-Universit\"at, Friedrich-Hund-Platz 1, 37077 G\"ottingen, Germany 
           \and Zentrum f\"ur Astronomie der Universt\"at Heidelberg, Landessternwarte, K\"onigstuhl 12, D-69117 Heidelberg, Germany 
           \and Instituto de Astrof\'isica de Andaluc\'ia (IAA-CSIC), Glorieta de la Astronom\'ia s/n, E-18008 Granada, Spain 
           \and Centre for Exoplanets and Habitability, University of Warwick, Gibbet Hill Road, Coventry, CV4 7AL, UK 
           \and Department of Physics, University of Warwick, Gibbet Hill Road, Coventry, CV4 7AL, UK 
           \and European Southern Observatory, Alonso de C\'{o}rdova 3107, Vitacura, Casilla 19001, Santiago 19, Chile 
           \and Instituto de Astrof\'{\i}sica e Ci\^encias do Espa\c{c}o, CAUP, Universidade do Porto, Rua das Estrelas, PT4150-762 Porto, Portugal 
           \and Departamento de F\'isica, Universidade Federal do Rio Grande do Norte, 59078-970 Natal, RN, Brazil 
           \and Geneva Observatory, University of Geneva, Chemin des Mailettes 51, 1290 Versoix, Switzerland 
           \and Center for Space and Habitability, Universit\"at Bern, Gesellschaftsstrasse 6, 3012 Bern, Switzerland 
         }

\date{Received November 07, 2019; accepted November 08, 2019}

\abstract{We report the discovery of a new planetary system with three transiting planets, one super-Earth and two sub-Neptunes, that orbit EPIC\,249893012, a G8\,IV-V evolved star ($M_\star$\,=\,1.05\,$\pm$\,0.05\,$M_\odot$, $R_\star$\,=\,1.71\,$\pm$\,0.04\,$R_\odot$, $T_\mathrm{eff}$\,=5430\,$\pm$\,85\,K). The star is just leaving the main sequence. We combined \ktwo \ photometry with IRCS adaptive-optics imaging and HARPS, HARPS-N, and CARMENES high-precision radial velocity measurements to confirm the planetary system, determine the stellar parameters, and measure radii, masses, and densities of the three planets. With an orbital period of $3.5949^{+0.0007}_{-0.0007}$ days, a mass of $8.75^{+1.09}_{-1.08}\ M_{\oplus}$ , and a radius of $1.95^{+0.09}_{-0.08}\ R_{\oplus}$, the inner planet b is compatible with nickel-iron core and a silicate mantle ($\rho_b= 6.39^{+1.19}_{-1.04}$ g cm$^{-3}$). Planets c and d with orbital periods of $15.624^{+0.001}_{-0.001}$ and $35.747^{+0.005}_{-0.005}$ days, respectively, have masses and radii of $14.67^{+1,84}_{-1.89}\ M_{\oplus}$ and $3.67^{+0.17}_{-0.14}\ R_{\oplus}$ and $10.18^{+2.46}_{-2.42}\ M_{\oplus}$ and $3.94^{+0.13}_{-0.12}\ R_{\oplus}$, respectively, yielding a mean density of $1.62^{+0.30}_{-0.29}$ and $0.91^{+0.25}_{-0.23}$ g cm$^{-3}$, respectively. The radius of planet b lies in the transition region between rocky and gaseous planets, but its density is consistent with a rocky composition. Its semimajor axis and the corresponding photoevaporation levels to which the planet has been exposed might explain its measured density today. In contrast, the densities and semimajor axes of planets c and d suggest a very thick atmosphere. The singularity of this system, which orbits a slightly evolved star that is just leaving the main sequence, makes it a good candidate for a deeper study from a dynamical point of view.}


\maketitle

\section{Introduction}

With the advent of space-based transit-search missions, the detection and characterization of exoplanets have undergone a fast-paced revolution. First \corot\ \citep{Auvergne2010} and then \kepler\ \citep{Borucki2010} marked a major leap forward in understanding the diversity of planets in our Galaxy. With the failure of its second reaction wheel, \kepler\ embarked on an extended mission, named \textit{K2} \citep{Howell2014}, which surveyed different stellar fields located along the ecliptic. Their high-precision photometry has allowed the \kepler\ and \textit{K2} missions to dramatically extended the parameter space of exoplanets, bringing the transit detection threshold down to the Earth-sized regime. The Transiting Exoplanet Survey Satellite \citep[\tess;][]{Ricker2015} is currently extending this search to cover almost the entire sky; it mainly focuses on bright stars (V\,$<$\,11).

Although super-Earths ($R_\mathrm{p}$\,$\simeq$\,$1-2$\,$R_{\oplus}$, $M_\mathrm{p}$\,$\simeq$\,$1-10$\,$M_{\oplus}$) and Neptune-sized planets ($R_\mathrm{p}$\,$\simeq$\,$2-4$\,$\ R_{\oplus}$, $M_\mathrm{p}$\,$\simeq$\,$10-40$\,$\, M_{\oplus}$) are ubiquitous in our Galaxy  \citep[see, e.g.,][]{Marcy2014,Silburt2015,Hsu2019}, we still have much to learn about the formation and evolution processes of small planets. Observations have led to the discovery of peculiar patterns in the parameter space of small exoplanets \citep{Winn2018}. The radius--period diagram shows a dearth of short-period Neptune-sized planets, the so-called Neptunian desert \citep{Mazeh2016, Owen2016}. Small planets tend to prefer radii of either $\sim$1.3\,$R_\oplus$ or $\sim$2.6\,$R_\oplus$, with a dearth of planets at $\sim$1.8\,$R_\oplus$, the so-called radius gap \citep{Fulton2017, Fulton2018}. Atmospheric erosion by high-energy stellar radiation (also known as photoevaporation) is believed to play a major role in shaping both the Neptunian desert and the bimodal distribution of planetary radii. Moreover, \cite{Armstrong2019} found a gap in the mass distribution of planets with a mass lower than $\sim$20\,$M_{\oplus}$ and periods shorter than 20 days, so far without any apparent physical explanation.

Understanding the formation and evolution of small planets requires precise and accurate measurements of their masses and radii. The \texttt{KESPRINT} consortium\footnote{\url{http://www.kesprint.science/}.} aims at confirming and characterizing planetary systems from the \ktwo\ mission \citep[see, e.g.,][]{Grziwa2016, Gandolfi2017,Jorge2019, Luque2019A&A...623A.114L, Palle2019}, and more recently, from the \tess\ mission \citep{Esposito2019, Gandolfi2018, Gandolfi2019}.

This paper is organized as follows: in Sect.\,\ref{sec:photo} we describe the \ktwo\ photometry together with the detection of the three transiting planets and a preliminary fit of their transit light curves. In Sect.\,\ref{sec:gnd} we describe our follow-up observations. The stellar fundamental parameters are provided in Sect.\,\ref{sec:stellar}. In Sect.\,\ref{sec:freq} we present the frequency analysis of the radial velocity measurements; the joint modeling is described in Sect.\,\ref{sec:joint}. Discussion and conclusions are given in Sect.\,\ref{sec:disc}.


\section{K2 photometry and detection}\label{sec:photo}
\begin{figure}
\centering
   \includegraphics[width=8.5cm]{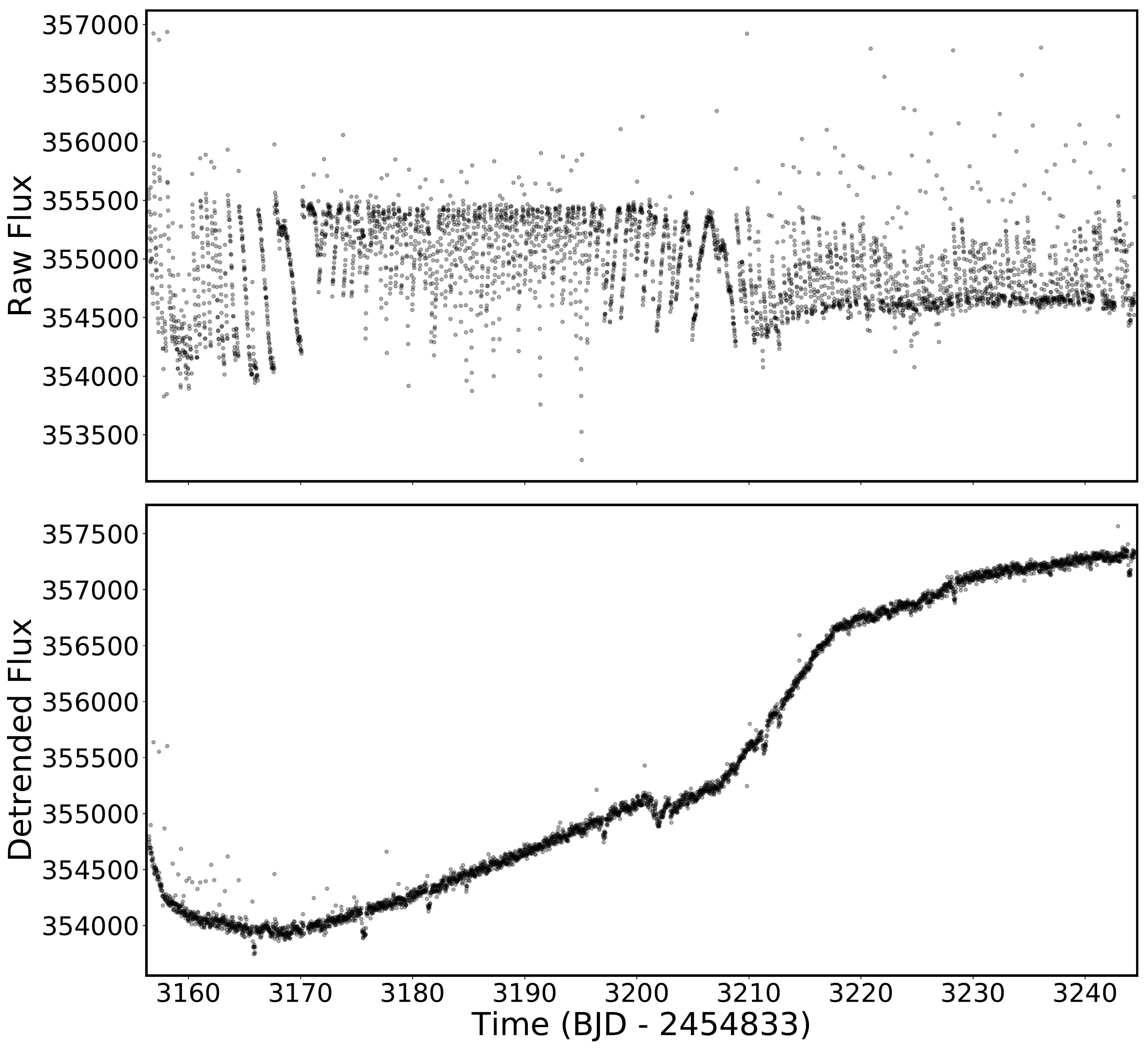}
     \caption{\ktwo\ light curves of \sname. The upper panel shows the raw light curve as extracted from the pixel data file in units of electrons per cadence. The lower panel shows the detrended light curve as obtained using our Everest-based pipeline. No stellar variability is detectable but the transit signals are clearly visible. }
     \label{fig:raw_lc}
\end{figure}

\sname\ was observed during \ktwo\ Campaign 15 of \ktwo\ as part of the \ktwo\ guest observer (GO) programs GO-15052 (PI: Stello D.) and GO-15021 (PI: Howard, A.\,W.). Campaign 15 lasted 88\,days, from 23 August 2017 to 20 November 2017, observing a patch of sky toward the constellations of Libra and Scorpius. During Campaign 15, the Sun emitted 27 M-class and four X-class flares and released several powerful coronal mass ejections (CMEs\footnote{\url{https://www.nasa.gov/feature/goddard/2017/september-2017s-intense-solar-activity-viewed-from-space}}). This affected the measured dark current levels for all \ktwo\ channels. Peak dark current emissions occurred around BJD 2458003.23, 2458007.85, and 2458009.00 (3170.23, 3174.85 and 3176.00, respectively, for the time reference value, BJD - 2454833, given in Fig.\,\ref{fig:raw_lc}).

\begin{figure}
\centering
   \includegraphics[width=7cm]{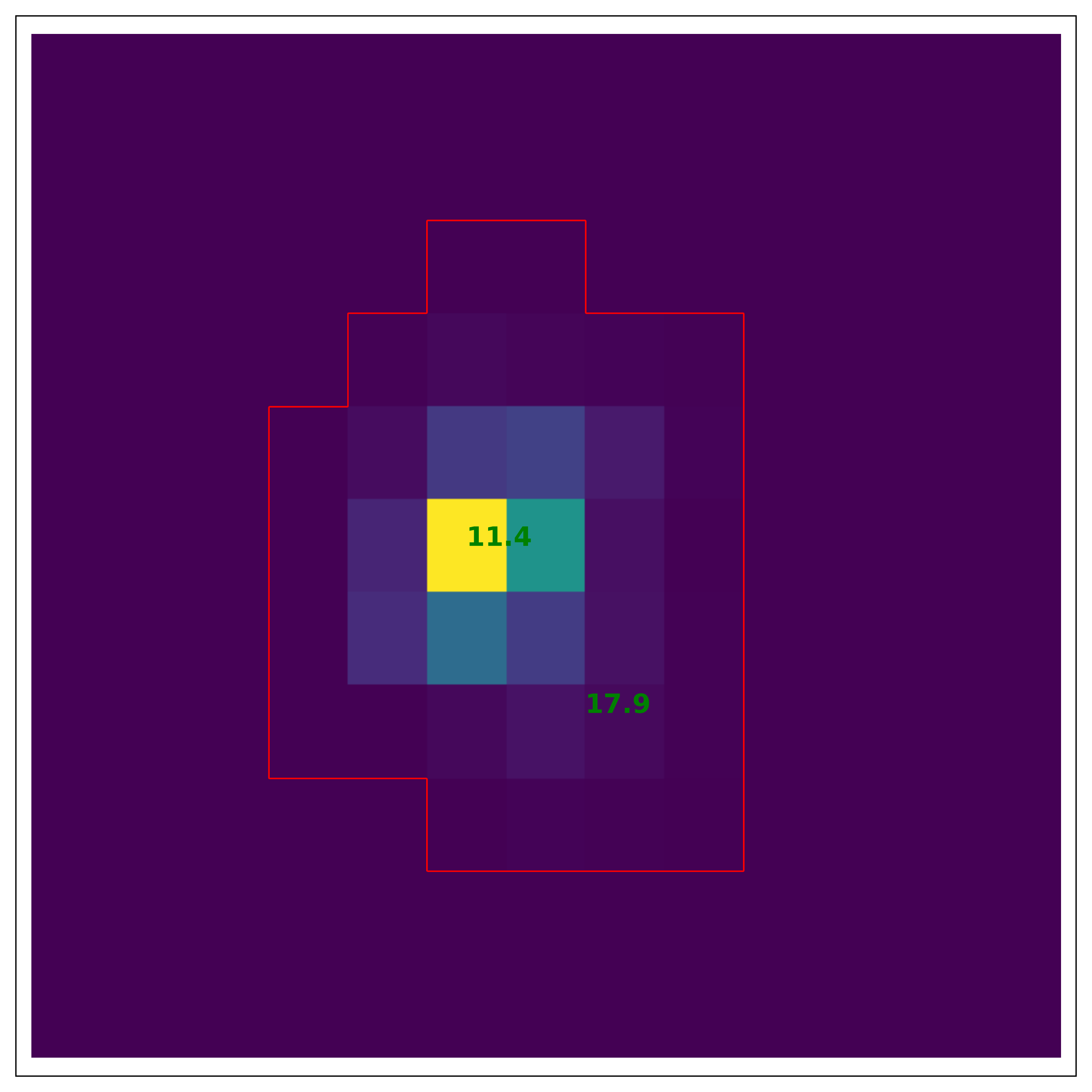}
     \caption{Customized \ktwo\ image of \sname. North is to the left and east at the bottom. The field of view is 43.78$\times$51.74\,arcsec (3.98\,arcsec per pixel). The red line marks the customized aperture for light-curve extraction, with a threshold of 1.7$\sigma$ above the background. Green annotations are the Kepler magnitude (retrieved from the RA and DEC from MAST) of \sname, and the source of contamination is identified in Fig.\ref{fig:ao-full_image}. }
     \label{fig:aper}
\end{figure}

We built the light curve of \sname\ from the target pixel file downloaded from the Mikulski Archive for Space Telescopes (MAST\footnote{\url{https://archive.stsci.edu/missions/k2/target\_pixel\_files/c15/249-800000/93000/ktwo249893012-c15\_lpd-targ.fits.gz}}). The pipeline used in this paper is based on the pixel level decorrelation (PLD) method that was initially developed by \citet{Deming2015} to correct the intra-pixel effects for {\it Warm Spitzer} data, and which was implemented in a modified and updated version of the \texttt{Everest}\footnote{\url{https://github.com/rodluger/everest}} pipeline \citep{Luger2018}. Our pipeline customizes different apertures for every single target by selecting the photocenter of the star and the nearest pixels, with a threshold of 1.7$\sigma$ above the previously calculated background (Fig.\,\ref{fig:aper}). After the aperture pixels were chosen, our pipeline extracted the raw light curve and removed all time cadences that were flagged as bad-quality data. The pipeline applies PLD to the data up to third order to perform robust flat-fielding corrections, which avoids us having to solve for correlations on stellar positions. It also uses a second step of Gaussian processes (GP), which separates astrophysical and instrumental variability, to compute the covariance matrix as described in \cite{Luger2018}. The raw and the final detrended light curves are plotted in Fig.\,\ref{fig:raw_lc}. Our pipeline, which is based on EVEREST, tends to introduce long-term modulation, masking low-frequency signals such as the stellar variability that is uncovered with the frequency analysis of the radial velocity data in section \ref{sec:freq}.

\begin{figure}
\centering
   \includegraphics[width=7cm]{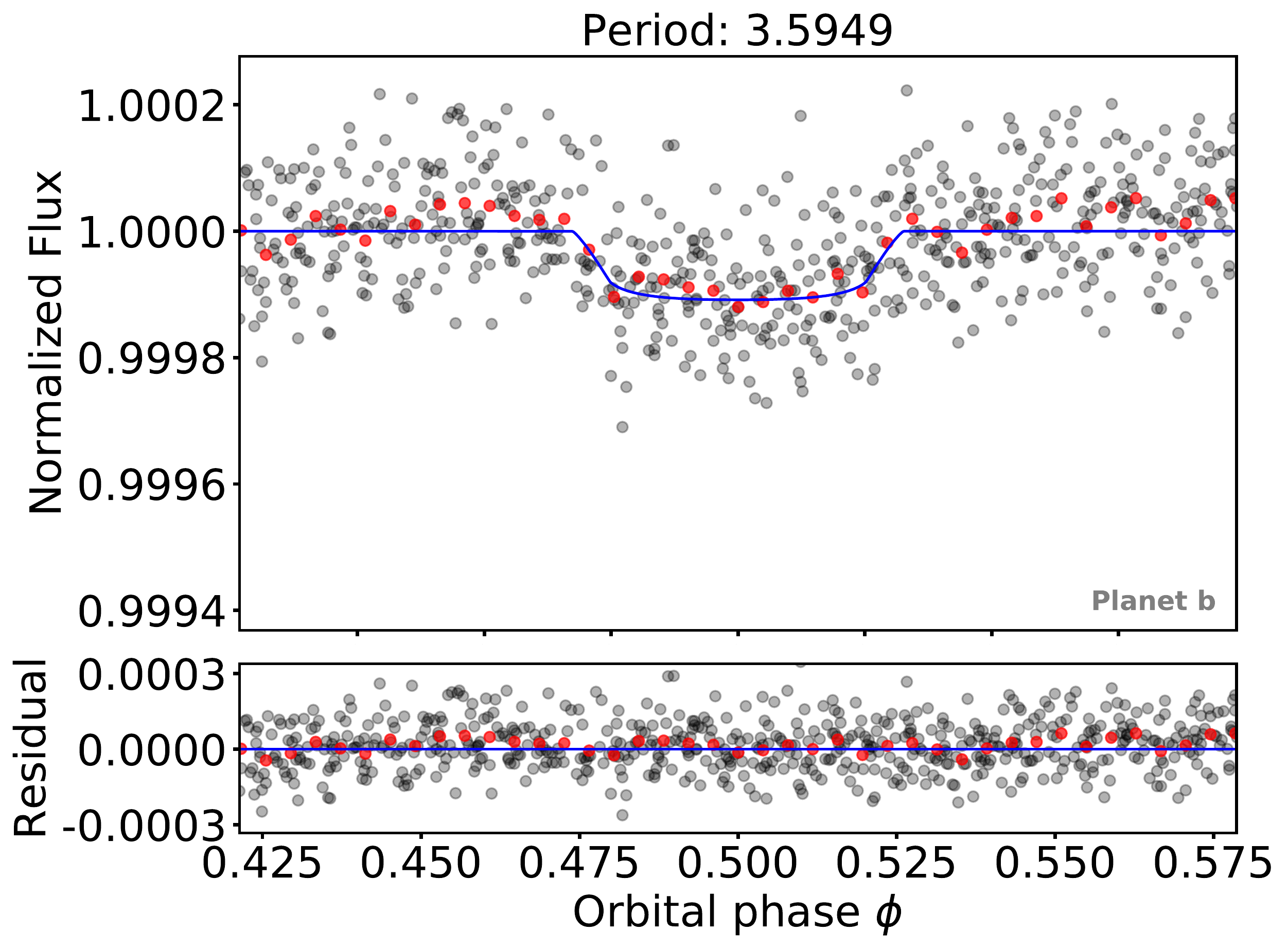}
   \includegraphics[width=7.1cm]{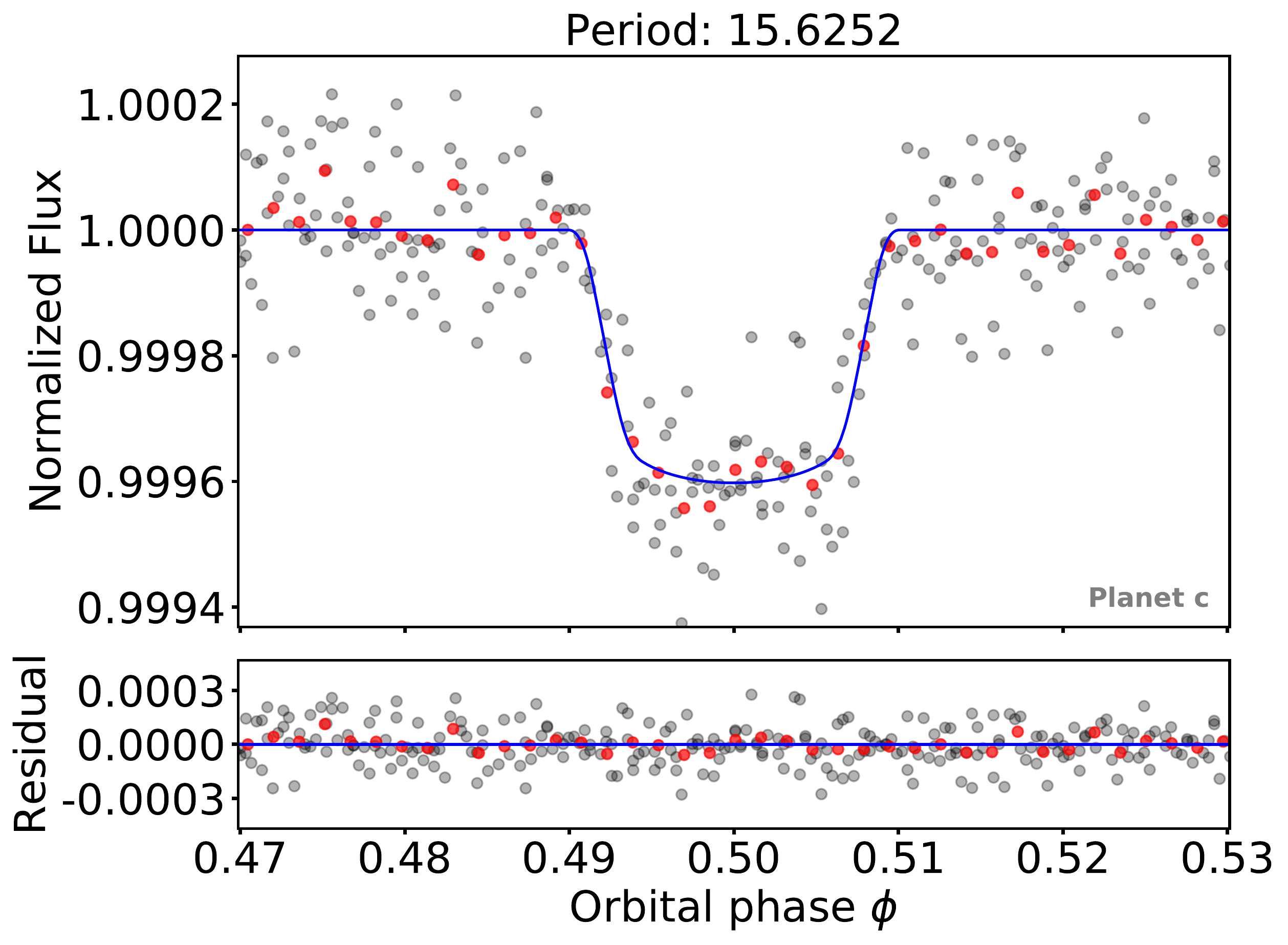}
   \includegraphics[width=7cm]{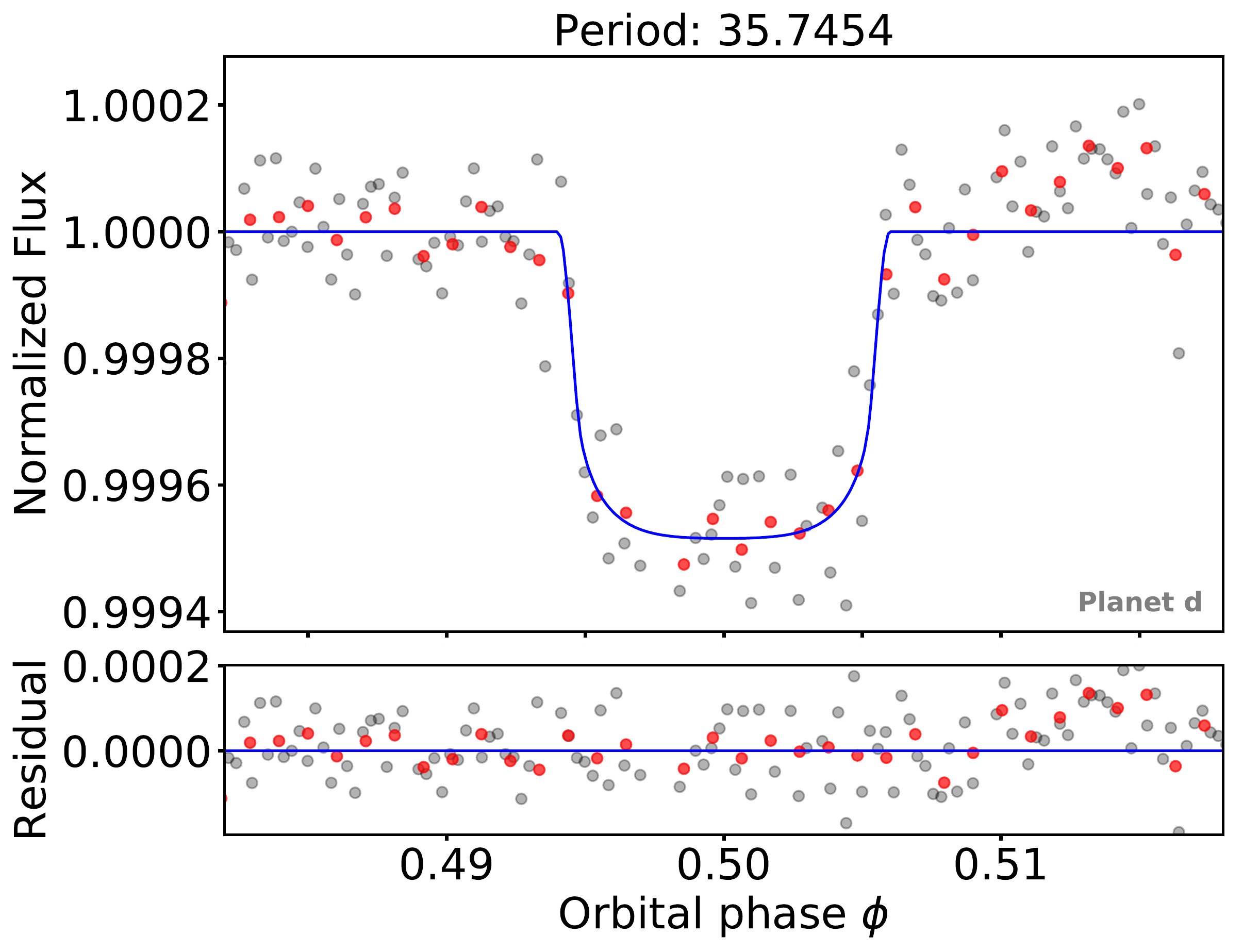}
     \caption{Phase-folded transit light curves of \sname\ b, c, and d (upper, middle, and lower panel, respectively). The black points mark the detrended \ktwo\ data. The red points mark the bins of 15 (top panel), 6 (middle panel), and 4 data points (bottom panel). The blue solid line represents the best-fit transit model for each planet. Residuals are shown in the lower panels of each transit light curve.}
     \label{fig:transits}
\end{figure}

We used a robust locally weighted regression method \citep{Cleveland1978}, with a fraction parameter of 0.04, to flatten the light curve. We also used this method with a lower fraction rate to interactively detect and remove outliers above 3$\sigma$ until no points were detected. We removed the data observed from the first three and a half days of the campaign, from BJD 2457989.44 to 2457993.0 (3156,44 to 3160.00 for the BJD - 2454833 time reference in Fig.\,\ref{fig:raw_lc}), because of a sharp trend at the beginning of the observation that is probably related to a thermal anomaly. We finally flattened the original light curve by dividing it by the model.

We searched the flattened light curve for transits using the box-fitting least squares (BLS) algorithm \citep{Kovac2002}. When a planetary signal was detected in the power spectrum, we fit a transit model using the \textsc{python} package \textit{batman} \citep{Laura2015}. We divided the transit model by the flattened light curve and again applied the BLS algorithm to find the next planetary signal, until no significant peak was found in the power spectrum.

We found three planetary signals in the \sname\ light curve, with periods of 3.59, 15.63, and 35.75 days and depths of 108.7, 402.3 and 484.3 ppm. The period ratios are 1:4.34:9.94, out of resonance, except for signals b and d with a ratio close to 1:10. Figure~\ref{fig:transits} shows the phase-folded light curve for each transit signal and the best-fit model.


\section{Ground-based follow-up observations}\label{sec:gnd}

\subsection{High-resolution imaging}
\label{subsec:hri}

\begin{figure}
\centering
   \includegraphics[width=8cm]{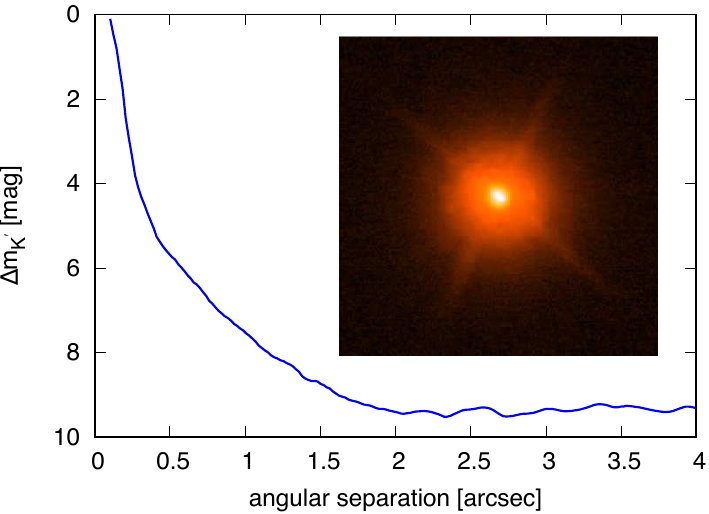}
     \caption{5\,$\sigma$ contrast curve against angular separation from \sname, based on the IRCS AO imaging. The inset exhibits \sname\ $4^{\prime\prime} \times 4^{\prime\prime}$ image. }
     \label{fig:ao}
\end{figure}

On 14 July 2019, we performed adaptive-optics (AO) imaging for \sname\ with the InfraRed Camera and Spectrograph \citep[IRCS:][]{2000SPIE.4008.1056K} on the Subaru 8.2m telescope to search for faint nearby sources that might contaminate the \ktwo\ photometry. Adopting the target star itself as a natural guide for AO, we imaged the target in the $K^\prime$ band with a five-point dithering. We obtained both short-exposure (unsaturated; $0.5s \times 3$ coaddition for each dithering position) and long-exposure (mildly saturated; $2.0s \times 3$ coaddition for each) frames of the target for absolute flux calibration and for inspecting nearby faint sources, respectively. We reduced the IRCS data following \citet{2016ApJ...820...41H}, and obtained the median-combined images for unsaturated and saturated frames, respectively. Based on the unsaturated image, we estimated the target full width at half-maximum (FWHM) to be $0\farcs115$. In order to estimate the detection limit of nearby faint companions around \sname, we computed the 5\,$\sigma$ contrast as a function of angular separation based on the flux scatter in each small annulus from the saturated target. Figure~\ref{fig:ao} plots the 5\,$\sigma$ contrast along with the $4^{\prime\prime} \times 4^{\prime\prime}$ target image in the inset. Our AO imaging achieved approximately $\Delta K^\prime=8$ mag at $1^{\prime\prime}$ from \sname.

\begin{figure}
\centerline{\includegraphics[angle=0,width=8cm]{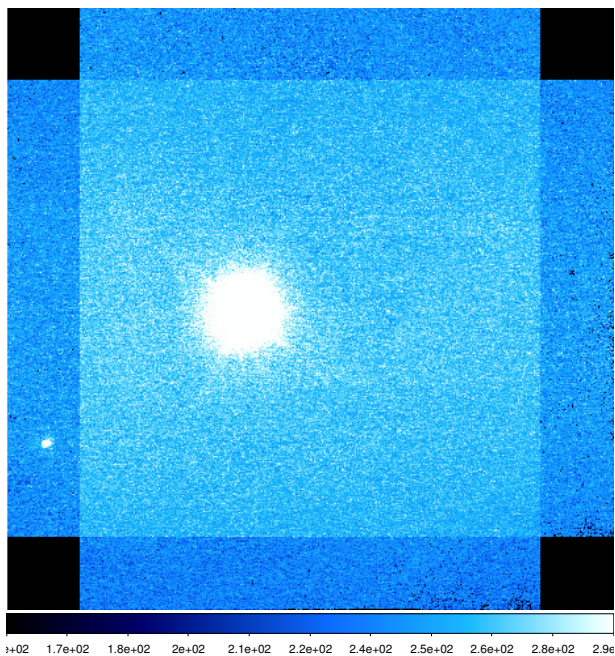}}
\caption{Adaptive-optics image of \sname\ obtained with the Subaru/IRCS instrument. North is up and east is to the left. Field of view of is 21\arcsec\ in both directions (pixel scale of $0\farcs02$/pix). Because this image was created after median-combining the aligned frames, background levels as well as flux scatters in the corners are different from those of the central part of the detector.
\label{fig:ao-full_image}}
\end{figure}

Visual inspection of the saturated image suggests no nearby companion within $5^{\prime\prime}$ from \sname, but it exhibits a faint source separated by $8\farcs3$ in the southeast (Fig.~\ref{fig:ao-full_image}), that is, inside the aperture for light-curve extraction on the \ktwo\ image (see Fig.~\ref{fig:aper}). Checking the \gaia\ DR2 catalog, we found that this faint source is the \gaia~DR2~6259260177825579136 star with $G=17.9$ mag \citep[\gaia~G magnitude defined in][]{Evans2018}; further information of this source is provided in Table~\ref{table-03}. The transit signal with depth of 100\,ppm on \sname\ ($K_{p} = 11.364$~mag) may be mimicked by an equal-mass eclipsing binary that is 9.29 magnitude fainter, that is, with a \kepler\ magnitude\footnote{\kepler~magnitude defined in \citet{Brown2011}.} of 20.65. Taking into account the close similarity between the \gaia\ and \kepler\ bandpasses, we therefore cannot exclude \gaia~DR2~6259260177825579136 as a source of a false positive for one of the three transit signals (see Sect.~\ref{subsec:faint_ao_companion}).

\subsection{High-resolution spectroscopy}

We collected 74 high-resolution (R$\approx$115\,000) spectra of \sname\ using the High Accuracy Radial velocity Planet Searcher (HARPS) spectrograph \citep{Mayor2003} mounted at the ESO-3.6m telescope of the La Silla observatory (Chile). The observations were carried out between April 2018 and August 2019 as part of our radial velocity (RV) follow-up of \ktwo\ and \tess\ planets conducted with the HARPS spectrograph (observing programs 0101.C-0829 and 1102.C-0923; PI: Gandolfi) and under program 60.A-9700 (technical time). We reduced the data using the HARPS Data Reduction Software (DRS) and extracted the Doppler measurements by cross-correlating the Echelle spectra with a G2 numerical mask \citep{Baranne1996, Pepe2002, Lovis2007}. The DRS also provides the user with the FWHM and the bisector inverse slope (BIS) of the cross-correlation function (CCF), as well as with the Ca {\sc ii} H\,\&\,K lines activity indicator\footnote{Extracted assuming a color index B-V\,=\,0.778.} log\,R$^\prime_\mathrm{HK}$.

Between April 2018 and March 2019, we also secured 11 high-resolution (R$\approx$115 000) spectra with the HARPS-N spectrograph \citep{Cosentino2012} mounted at the 3.58m Telescopio Nazionale Galileo at Roque de Los Muchachos observatory (La Palma, Spain), as part of the observing programs CAT18A\_130, CAT18B\_93 (PI: Nowak), and A37TAC\_37 (PI: Gandolfi). The data reduction, as well as the extraction of the RV measurements and activity- and line-profile indicators follows the same procedure as for the HARPS spectra.

Between 6 May 2018 and 21 June 2018, we also collected 25 spectra of \sname\ with the Calar Alto high-Resolution search for M dwarfs with Exoearths with Near-infrared and optical Échelle Spectrographs (CARMENES) instrument \citep{CARMENES,CARMENES18}, installed at the 3.5m telescope of Calar Alto Observatory in Spain (observing program S18-3.5-021 -- PI: Pall{\'e}.). The instrument consists of a visual (VIS, $\SI{0.52}- \SI{0.96}{\micro\metre}$) and a near-infrared (NIR, $\SI{0.96} - \SI{1.71}{\micro\metre}$) channel yielding spectra at a resolution of $R\,\approx\,94\,600$ and $R\approx80\,400$, respectively. Like \citet{Luque2019A&A...623A.114L}, we only used the VIS observations to extract the RV measurements because the spectral type of \sname\ is solar like. We computed the CCF using a weighted mask constructed from the coadded CARMENES VIS spectra of \sname\ and determined the RV, FWHM, and the BIS measurements by fitting a Gaussian function to the final CCF following the method described in \citet{Reiners18}.

Tables~\ref{table:RV-H}, \ref{table:RV-HN}, and \ref{table:RV-C} list the HARPS, HARPS-N, and CARMENES Doppler measurements and their uncertainties, along with the BIS and FWHM of the CCF, the exposure time, the signal-to-noise ratio (S/N) per pixel at 5500\,\AA\ for HARPS and HARPS-N, and at 5340\,\AA\ for CARMENES, and for HARPS and HARPS-N alone, the Ca {\sc ii} H \& K activity index log\,R$^\prime_\mathrm{HK}$.


\begin{table}[t]
\caption{Relative properties of the nearby star to \sname\ detected  with
the Subaru/IRCS.
\label{table-03}}
\begin{center}
\begin{tabular}{lc}
\hline
\hline
\noalign{\smallskip}
Parameter                      & close-in star\\
\noalign{\smallskip}
\hline
\noalign{\smallskip}
Separation (\arcsec)           & $  8.30 \pm 0.03$\\
Position Angle (deg)           & $ 124.12   \pm 0.10$\\
$\Delta m_{K^\prime}$ (mag)    & $  6.697 \pm 0.023 $\\
$\Delta F_{K^\prime}$ relative flux   & $  (2.095 \pm 0.044) \times 10^{-3}$\\
\noalign{\smallskip}
\hline
\end{tabular}
\end{center}
\end{table}

\section{Stellar properties}\label{sec:stellar}

\begin{table}[t!]
\centering
\caption{Equatorial coordinates, main identifiers, optical and infrared magnitudes, proper motion, parallax, distance, spectroscopic parameters, interstellar extinction and fundamental parameters of \sname.}
\label{table:stellar_params}
\begin{threeparttable}
\begin{tabular}{lrr}
\hline
\hline
\noalign{\smallskip}
Parameter & Value & Source\\
\noalign{\smallskip}
\hline
\noalign{\smallskip}
\multicolumn{3}{c}{\emph{Equatorial Coordinates and Main Identifiers}}\\
\noalign{\smallskip}
RA$_\mathrm{J2000.0}$ (hh:mm:ss) & 15:12:59.57 & \gaia \tnote{a}\\
DEC$_\mathrm{J2000.0}$ (dd:mm:ss) & $-$16:43:28.73 & \gaia \tnote{a}\\
\gaia \ ID & {\small 6259263137059042048} & \gaia \tnote{a}\\
2MASS ID & {\small J15125956-1643282} & 2MASS\tnote{b}\\
TYC ID & 6170-95-1 & TYCHO2\tnote{c}\\
\noalign{\smallskip}
\hline
\noalign{\smallskip}
\multicolumn{3}{c}{\emph{Optical and Near-Infrared Magnitudes}}\\
\noalign{\smallskip}
$K_{p}$ (mag) & 11.364 & \ktwo \tnote{d}\\
$B_{J}$ (mag) & 12.335 $\pm$ 0.240 & \ktwo \tnote{d}\\
$V_{J}$ (mag) & 11.428 $\pm$ 0.121 & \ktwo \tnote{d}\\
$G$ (mag) & 11.4019 $\pm$ 0.0005 &   \gaia \tnote{a}\\
$g$ (mag) & 11.911 $\pm$ 0.010 & \ktwo \tnote{d}\\
$r$ (mag) & 11.370 $\pm$ 0.020 & \ktwo \tnote{d}\\
$i$ (mag) & 11.130 $\pm$ 0.030 & \ktwo \tnote{d}\\
$J$ (mag) & 10.216 $\pm$ 0.026 & \ktwo \tnote{d}\\
$H$ (mag) & 9.800 $\pm$ 0.023 & \ktwo \tnote{d}\\
$K$ (mag) & 9.714 $\pm$ 0.023 & \ktwo \tnote{d}\\
\noalign{\smallskip}
\hline
\noalign{\smallskip}
\multicolumn{3}{c}{\emph{Space Motion and Distance}}\\
\noalign{\smallskip}
PM$_\mathrm{RA}$ (mas \ yr$^{-1}$) & 13.55 $\pm$ 0.07 &   \gaia \tnote{a}\\
PM$_\mathrm{DEC}$ (mas \ yr$^{-1}$) & $-$34.29 $\pm$ 0.06 &   \gaia \tnote{a}\\
Parallax (mas) & 3.08 $\pm$ 0.04 & \gaia \tnote{a}\\
Distance (pc) & 324.7 $\pm$ 4.2 &   \gaia \tnote{a}\\
\noalign{\smallskip}
\hline
\noalign{\smallskip}
\multicolumn{3}{c}{\emph{Spectroscopic Parameters and Interstellar Extinction}}\\
\noalign{\smallskip}
Spectral type & G8 IV/V & This work\\
\teff \ (K) & 5430 $\pm$ 85 & This work\\
\logg \ (cgs) & 3.99 $\pm$ 0.03 & This work\\
\feh \ (dex) & 0.20 $\pm$ 0.05 & This work\\
\mgh \ (dex) & 0.28 $\pm$ 0.05 & This work\\
\nah \ (dex) & 0.25 $\pm$ 0.05 & This work\\
\cah \ (dex) & 0.18 $\pm$ 0.05 & This work\\
\vmac \ (km s$^{-1}$) & 3.5 $\pm$ 0.4& This work\\
\vmic \ (km s$^{-1}$) & 0.9 $\pm$ 0.1 & This work\\
\vsini \ (km s$^{-1}$) & 2.1 $\pm$ 0.5 & This work\\
$A_\mathrm{V}$ (mag) & 0.19 $\pm$ 0.02& This work\\
\noalign{\smallskip}
\hline
\noalign{\smallskip}
\multicolumn{3}{c}{\emph{Stellar Fundamental Parameters}}\\
\noalign{\smallskip}
\mstar ($M_{\odot}$) & 1.05 $\pm$ 0.05 & This work\\
\rstar ($R_{\odot}$) & 1.71 $\pm$ 0.04 & This work\\
                     & 1.81$^{+0.11}_{-0.27}$& \gaia \tnote{a}\\
\noalign{\smallskip}
$L_{\star}$ ($L_{\odot}$) & 2.26$^{+0.04}_{-0.05}$& \gaia \tnote{a}\\
\noalign{\smallskip}
$\rho_{\star}$ (\gc) & 0.298$^{+0.026}_{-0.023}$& This work\\
\noalign{\smallskip}
Age (Gyr)   & 9.0$^{+0.5}_{-0.6}$& This work\\
\noalign{\smallskip}
\hline
\end{tabular}
\begin{tablenotes}
\item[a]  \gaia \ DR2 \citep{gaiadr22018}.
\item[b] 2MASS Catalog \citep{Skrutskie2006}.
\item[c] TYCHO2 Catalog \citep{Hog2000}.
\item[d] ExoFOP\footnote{\url{https://exofop.ipac.caltech.edu/k2/edit_target.php?id=249893012}}.
\end{tablenotes}
\end{threeparttable}
\end{table}

\subsection{Photospheric parameters}\label{sec:pho_para}
We extracted the spectroscopic parameters of the host star from the co-added HARPS spectrum -- which has a S/N ratio per pixel of S/N=270 at 5500\,\AA\ -- using two publicly available packages, as described in the following paragraphs.

We first used the package Spectroscopy Made Easy (\texttt{SME}), version 5.22, \citep{Valenti1996, Valenti2005, Piskunov2017}. \texttt{SME} calculates the equation of state, the line and continuum opacities, and the radiative transfer over the stellar surface with the help of a library of stellar models. A chi-square minimization procedure is then used to extract spectroscopy parameters, that is, the effective temperature \teff, the surface gravity \logg, the metal content, the micro \vmic\ and macro \vmac\ turbulent velocities, and the projected-rotational velocity \vsini, as described in \cite{Malcolm2017} and \cite{Persson2018, Persson2019}. When any of the parameters listed above an be determined with another method and it can be held fixed during the iterative procedure, this improves the determination of the remaining parameters. The turbulent velocities can typically be obtained as soon as the \teff\ is derived and/or can be inferred from empirical equations such as those of \citet{Bruntt2010} and \citet{Doyle2014}. In the case of EPIC\,249893012, we iteratively determined \teff\  by fitting the wings of the Balmer lines and then proceeded to obtain the other parameters. We selected the grid of \texttt{ATLAS12} models \citep{Kurucz2013} as the basis for our analysis. After obtaining the relevant abundances of metals, \logg\ was obtained by fitting the spectral lines of \mgi and \cai and checking by finally analyzing the Na\,{\sc i} doublet. The values for each parameter can be found in the Table \ref{table:stellar_params}. The result indicates a somewhat evolved solar-type star with a \logg \ of 3.8-3.9. Atomic and molecular parameters needed for the analysis were downloaded from the VALD database\footnote{\url{http://vald.astro.uu.se}.} \citep{Ryabchikova2015}.

We also used the package \texttt{specmatch-emp} \citep{Yee2017}, which uses a library of $\sim$400 high-resolution template spectra of well-characterized FGKM stars obtained with the HIRES spectrograph on the Keck telescope. We used a custom algorithm to put our HARPS spectrum into the same format as the HIRES spectra \citep{Hirano2018}, which was then compared to the spectra within the library to find the best match. \texttt{specmatch-emp} provides the effective temperature \teff\ and iron content \feh, along with the stellar radius, \rstar. We found values for \teff \ and \feh \ within 1\,$\sigma$ of the \texttt{SME}-derived values, as well as a stellar radius of \rstar\,=\,1.4 $\pm$ 0.2 \Rsun, which is consistent with the radius derived in Sect.\,\ref{subsec:mass}. The spectroscopic parameters of \sname\ imply a spectral type and luminosity class of G8 IV/V \citep{Cox2000, Gray2008}.

\subsection{Stellar mass, radius, and age} \label{subsec:mass}

Our data enable the measurement of the planetary fundamental parameters, most notably, the planetary radius, mass, and mean density. However, the stellar parameters are dependent on the properties of the host star. In order to extract the planetary properties and evaluate the evolutionary status of the planet, we need to derive the physical stellar parameters such as \mstar, \rstar, and age (assumed to be the same as that of the planet) using the spectral data.

We began by applying the spectroscopic parameters of Sect.~\ref{sec:pho_para} to the \cite{Torres2010a} empirical relation and derived preliminary estimates of the a stellar mass (1.3\,$\pm$\,0.1 \Msun) and radius (2.3\,$\pm$\,0.5\,\Rsun). In order to improve the precision, we used the \textit{Gaia} parallax \citep{gaiadr22018} along with the magnitudes listed in Table~\ref{table:stellar_params} and estimated the interstellar extinction along the line of sight to the star in two ways. The first method fits the spectral energy distribution (SED) using low-resolution synthetic spectra, as described in \cite{Gandolfi2008}, and gives an extinction of $A_\mathrm{V}$\,=\,0.25\,$\pm$\,0.08. The second method uses a 3D galactic dust map \citep{Green2018} to provide the extinction as $A_\mathrm{V}$\,= 0.19\,$\pm$\,0.02. The two methods are consistent within the uncertainties. We used the bolometric correction $BC_\mathrm{V}$ derived using the \cite{Torres2010b} corrections to the empirical equation of \cite{Flower1996} to derive the radius of the star as 1.67\,$\pm$\,0.09 \Rsun. We confirmed this value through the calculation of model tracks using the Bayesian \texttt{PARAM\,1.3} webtool \citep{Silva2006}\footnote{\url{http://stev.oapd.inaf.it/cgi-bin/param_1.3}}. Here we used the spectroscopic parameters, the dereddened Johnson visual magnitude $V_J$, and \textit{Gaia} parallax. \texttt{PARAM\,1.3} gives a stellar mass of 1.1\,$\pm$\,0.02\,\Msun\ with a radius consistent with the result derived above from the parallax and (dereddened) magnitude. The age is found to be about 7-8 Gyr.

Finally, we used the BAyesian STellar Algorithm \citep[\texttt{BASTA},][]{Silva2015} with a grid of MESA \citep[Modules for Experiments in Stellar Astrophysics,][]{Paxton2011} stellar models to perform a joint fit to the SED ($B_J, V_J, J, H, K, G$)  and spectroscopic parameters $T_{\rm eff}$, $\log g$, [Fe/H]. We adopted the extinction by \citet{Green2018} and corrected the parallax for the offset found in \citet{Stassun2018} while quadratically adding $0.1$\,mas to the uncertainty to account for systematics in the \textit{Gaia} DR2 data \citep{Luri2018}. We likewise corrected the \textit{Gaia} $G$-band magnitude following \citet{Casagrande2018} and adopted an uncertainty of 0.01\,mag. We found consistent values of $M_\star$\,=\,1.05 $\pm$ 0.05\,\Msun, $R_\star$\,=\,1.71 $\pm$ 0.04\,\Rsun and an age of 9.0$^{+0.5}_{-0.6}$ Gyr. We adopted these parameters for the analysis presented in the following sections.

The empirical and evolutionary model-dependent derivation of the stellar parameters, coupled with our spectroscopic parameters and \texttt{Gaia} parallax, confirm that \sname\ is a G-type star slightly more massive than the Sun in its first stages of evolution off the main sequence. Thus, it has a slightly lower \teff~than the Sun with a somewhat larger radius, as inferred by its significantly lower value of \logg.

\subsection{Faint AO companion}
\label{subsec:faint_ao_companion}

The faint star detected in the Subaru/IRCS AO image and identified as the Gaia DR2 6259260177825579136 star (see Section~\ref{subsec:hri}) cannot be excluded as a possible source of one of the transit signals detected in the \ktwo\ light curve of \sname. The parallax of Gaia DR2 6259260177825579136 ($\pi\,=\,0.3175\,\pm\,0.1573$~mas) and its proper motion ($\mathrm{PM_\mathrm{RA}}\,=\,-1.37 \pm 0.32\,\maspyr$ and $\mathrm{PM_\mathrm{DEC}}\,=\,-3.18\,\pm\,0.28\,\ \maspyr$) suggest that this is a distant background star. \cite{2018AJ....156...58B} determined the distance of Gaia DR2 6259260177825579136 to be $2.79^{+1.66}_{-0.87}$~kpc, that is, between 1.92 and 4.45~kpc. Using this value and the apparent magnitudes in the \gaia\ and K bandpasses, we calculated an absolute magnitude of Gaia DR2 6259260177825579136 of $M_{G} = 5.7^{+0.8}_{-1.0}$ and $M_{K} = 4.2^{+0.8}_{-1.0}$. Based on the \cite{2013ApJS..208....9P} and \cite{2012ApJ...746..154P} calibrations\footnote{Version 2019.3.22, available online at \url{http://www.pas.rochester.edu/~emamajek/EEM_dwarf_UBVIJHK_colors_Teff.txt}.} for absolute \gaia\ and K bandpasses, we estimated that Gaia DR2 6259260177825579136 is a G2--K8 dwarf star.

However, a false-positive scenario with the Gaia DR2 6259260177825579136 star as an equal-mass eclipsing binary is highly improbable for any of three transit signals of \sname. In the RVs of \sname\ we detect all three signals with the same periods as those found in the \ktwo\ light curve (Section~\ref{sec:photo}). None of these RV signals is visible in the chromospheric (log\,R$^\prime_\mathrm{HK}$) or photospheric activity indicators (FWHM and BIS of the CCFs; see Section ~\ref{sec:freq}). Therefore we conclude that they are Doppler signals induced by orbital motions of planets that transit \sname.

\section{Frequency analysis of the RV data}
\label{sec:freq}

In order to search for the Doppler reflex motion induced by the three transiting planets and unveil the presence of possible additional signals in our time-series Doppler data, we performed a frequency analysis of the RV measurements and their activity indicators. To this end, we used only the HARPS data taken in 2019. This allowed us to 1) avoid spurious peaks introduced by the one-year sampling and 2) avoid having to account for RV offsets between HARPS, HARPS-N, and CARMENES. The 60 HARPS RV measurements taken in 2019 cover a time baseline of about 171 d, translating into a spectral resolution of $171^{-1}$\,$\approx$\,0.006 d$^{-1}$.

\begin{figure}
\centering
   \includegraphics[width=\linewidth]{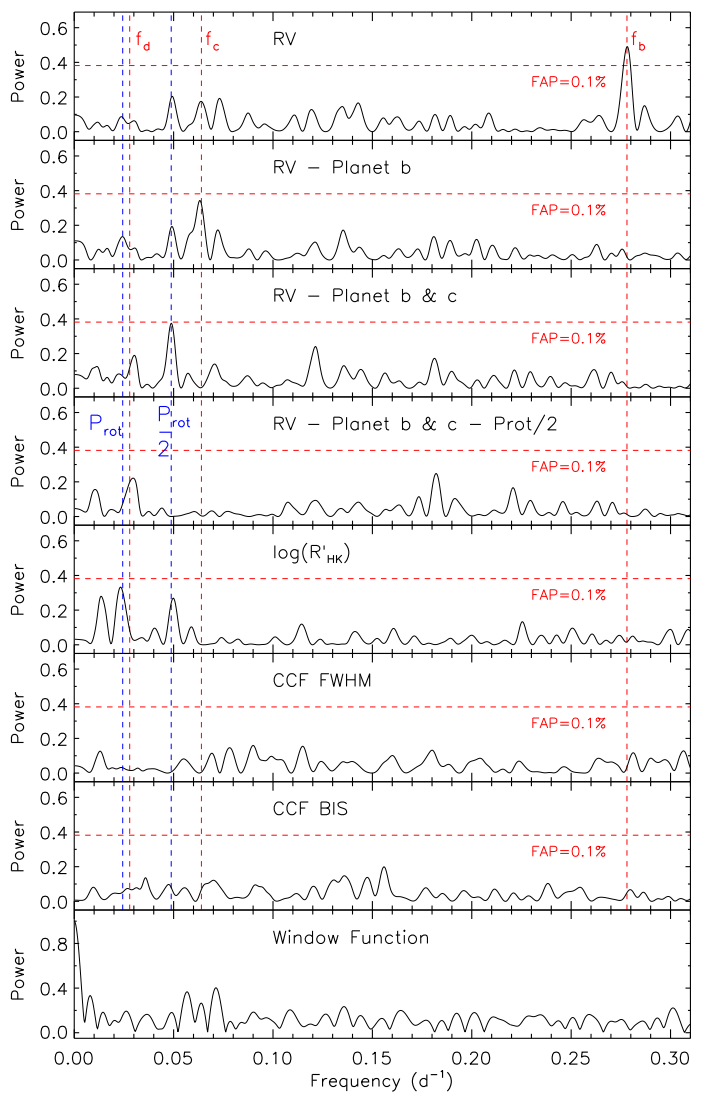}
     \caption{Generalized Lomb-Scargle periodogram of the 2019 HARPS measurements (upper panel) and RV residuals, following the subtraction of the Doppler signals of planet b (second panel), planets b and c (third panel), and planets b and c plus the 20.5 d signal (fourth panel). The periodogram of the Ca {\sc ii} H \& K lines activity indicator log\,R$^\prime_\mathrm{HK}$, of the CCF BIS and FWHM, and of the window function are shown in the last four panels. The horizontal dashed lines mark the 0.1\,\% FAP. The orbital frequencies of planets b, c, and d, as well as the stellar rotation frequency and its first harmonic are marked with vertical dashed lines.}
     \label{fig:periodogram}
\end{figure}

The upper panel of Figure~\ref{fig:periodogram} shows the generalized Lomb-Scargle periodogram \citep{Zechmeister2009} of the 2019 HARPS data. Following \citet{Kuerster1997}, the false-alarm probability (FAP) was assessed by computing the GLS periodogram of 10$^5$ mock time-series obtained by randomly shuffling the Doppler measurements, while keeping the time-stamps fixed. We found a significant peak at the orbital frequency of the inner transiting planet \sname\,b ($f_\mathrm{b}$\,=\,0.28\,d$^{-1}$, $P_\mathrm{b}\,=\,3.6$\,d), with an FAP\,$<$\,0.1\,\% over the frequency range 0.0--0.3\,d$^{-1}$. The \ktwo\ light curve provides prior knowledge of the possible presence of Doppler signals at three given frequencies, that is, the transiting frequencies. We therefore computed the probability that random data sets can result in a peak higher than the observed peak within a narrow spectral window centered around the transit frequency of the inner planet. To this aim we computed the FAP in a window centered around $f_\mathrm{b}$\,=\,0.28 d$^{-1}$ with a full width arbitrarily chosen to be six times the spectral resolution of the 2019 HARPS data (i.e., $6\,\times\,0.006$\,=\,0.036\,d$^{-1}$) and found an FAP\,$<$\,10$^{-5}$\,\%.

We computed the GLS periodogram of the RV residuals following the subtraction of the Doppler signal of \sname\,b. We fit the 2019 HARPS time series using the code \texttt{pyaneti} \citep[][see also Sect.~\ref{sec:joint}]{Barragan2018}, assuming that planet b has a circular orbit\footnote{We note that the orbits of the three planets are nearly circular and their eccentricities are consistent with zero (Sect.~\ref{sec:joint}).}, and kept both period and time of first transit fixed to the values derived from the \ktwo\ light curve, while allowing the RV semiamplitude to vary. The GLS periodogram of the RV residuals (Fig.~\ref{fig:periodogram}, second panel) shows a peak at the orbital frequency of \sname\,c ($f_\mathrm{c}$\,=\,0.06\,d$^{-1}$, $P_\mathrm{c}$\,=\,15.6~d) with an FAP\,$\approx$1\,\% over the frequency range 0.0--0.3\,d$^{-1}$. Analogously, the FAP in a narrow spectral window centered around $f_\mathrm{c}$\,=\,0.06\,d$^{-1}$ is $\sim$0.1\,\%.

We furthermore removed the RV signals of \sname\,b and c by performing a two-Keplerian joint fit to the HARPS data, assuming circular orbits and fixing periods and time of first transit to the \ktwo\ ephemeris. The GLS periodogram of the RV residuals, as obtained by subtracting the Doppler signals of the first two planets, displays a significant peak at $\sim$0.049\,c/d (FAP\,$<$\,0.1\,\%), corresponding to a period of about 20.5 days (Fig.~\ref{fig:periodogram}, third panel; see also next paragraph). We again subtracted this signal, along with the Doppler reflex motions of planets b and c, modeling the HARPS measurements with a sine curve and two circular Keplerian orbits. The periodogram of the RV residuals shows a peak close to the orbital frequency of the outer transiting planet \sname\,d ($f_\mathrm{d}$\,=\,0.028\,d$^{-1}$; Fig.~\ref{fig:periodogram}, fourth panel) whose FAP is, however, not significant (FAP\,$\approx$\,20\,\%) in the frequency domain 0.0--0.3\,d$^{-1}$. The probability that random time series can produce a peak higher than the observed peak in a narrow window centered around the frequency of the outer transiting planet is $\sim$1\,\%.

The nature of the 20.5-day signal remains to be determined. The panels 5-7 of Fig.~\ref{fig:periodogram} display the periodogram of the Ca {\sc ii} H\,\& K lines activity indicator (log\,R$^\prime_\mathrm{HK}$) and of the BIS and FWHM of the cross-correlation function, respectively. While the latter show no significant peaks at the orbital frequencies of the transiting planets or at 20.5 d ($\sim$0.049\,c/d), the periodogram of log\,R$^\prime_\mathrm{HK}$ displays a peak at 0.024\,d$^{-1}$ ($P\,=\,41$\,d), which is half the frequency (or twice the period) of the additional signal found in the RV residuals. The same periodogram also shows a peak at $\sim$0.049\,d$^{-1}$ (20.5\,d). Although none of the peaks seen in the periodogram of log\,R$^\prime_\mathrm{HK}$ has an FAP\,<\,0.1\,\%, we suspect that the rotation period of the star is $P_\mathrm{rot}$\,=\,41 d, and that the signal at 20.5 d is the first harmonic of the rotation period, which might arise from the presence of active regions at opposite longitudes carried around by stellar rotation. Assuming that the star is seen equator-on, the stellar radius of $R_\star$\,=\,1.71\,$\pm$0.04\,$R_\odot$ and the projected rotation velocity of \vsini\,=\,2.1\,$\pm$\,0.5\,\kms\ translate into a rotation period of 41\,$\pm$\,10 d, corroborating our interpretation.

\begin{figure*}
\centering
  \includegraphics[width=17.8cm]{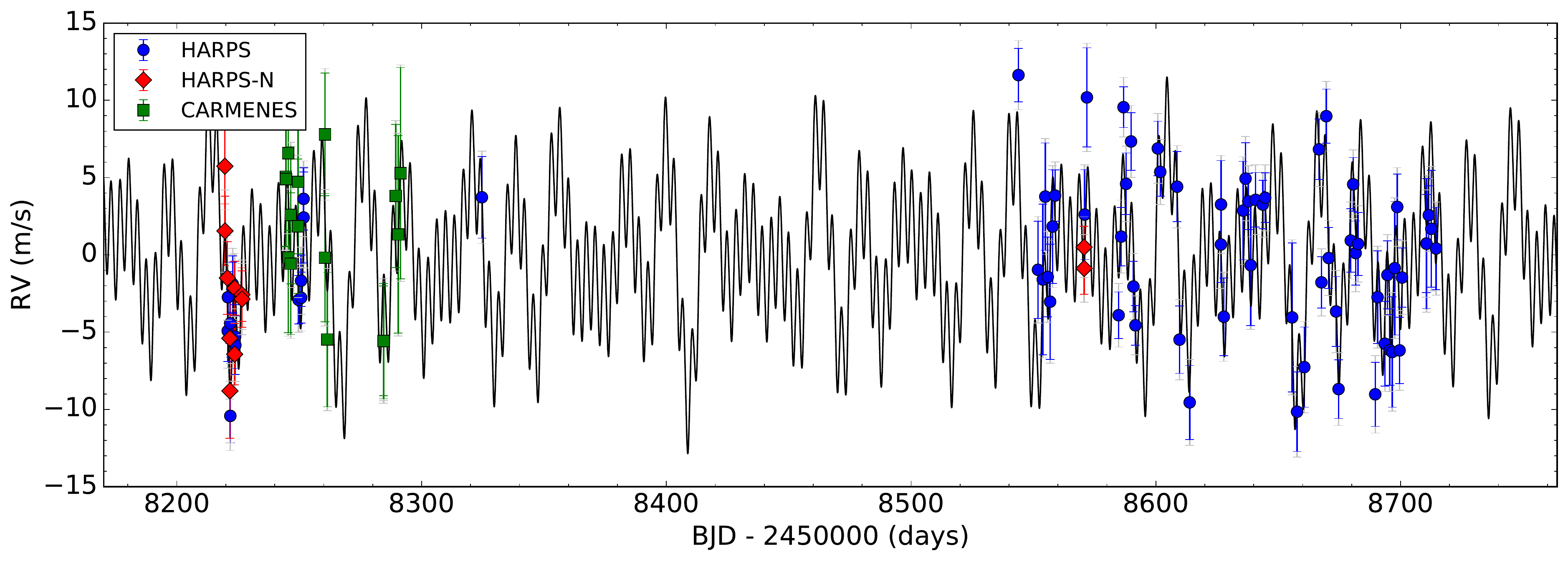}
  \includegraphics[width=9cm]{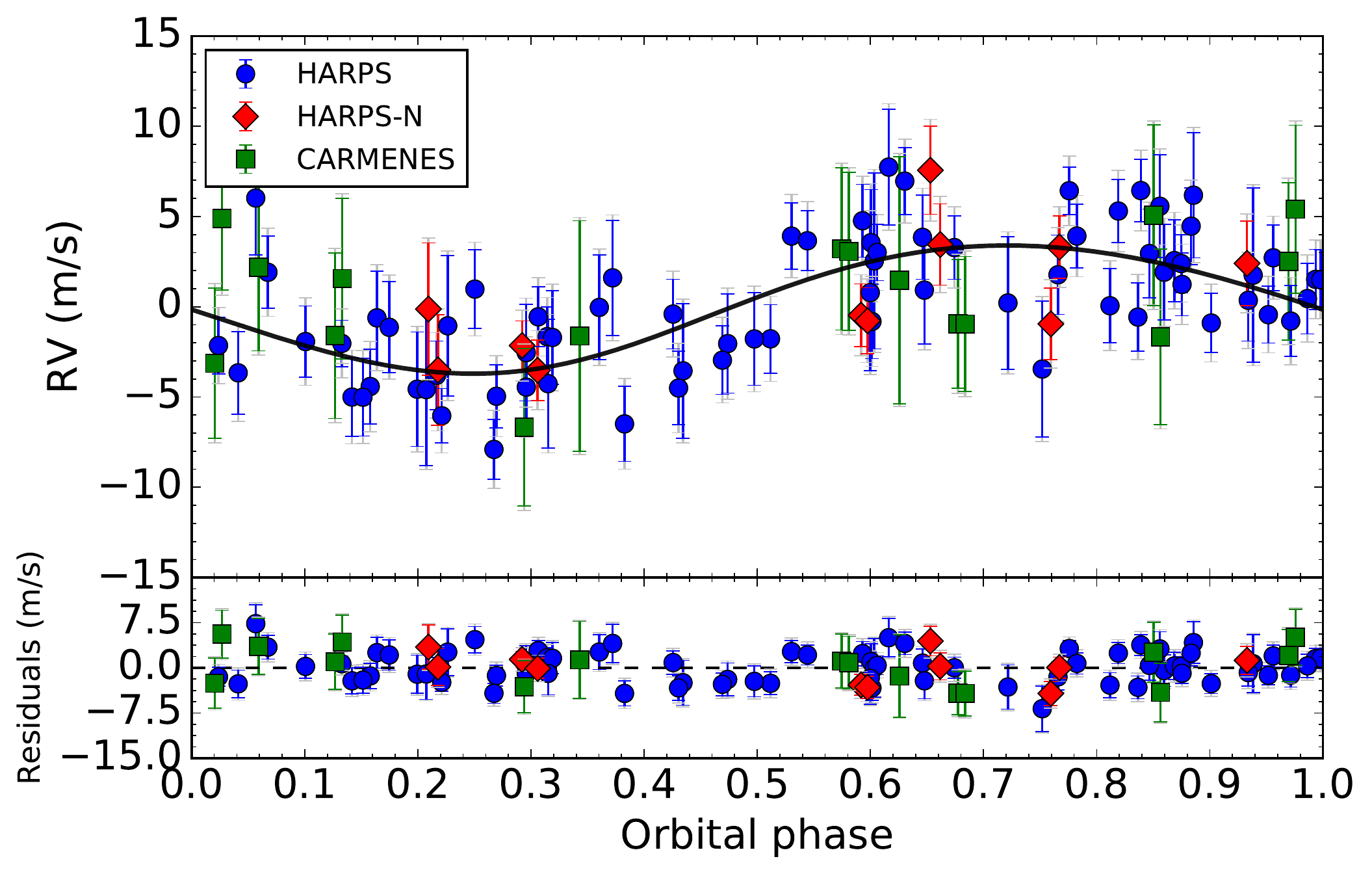}
  \includegraphics[width=9cm]{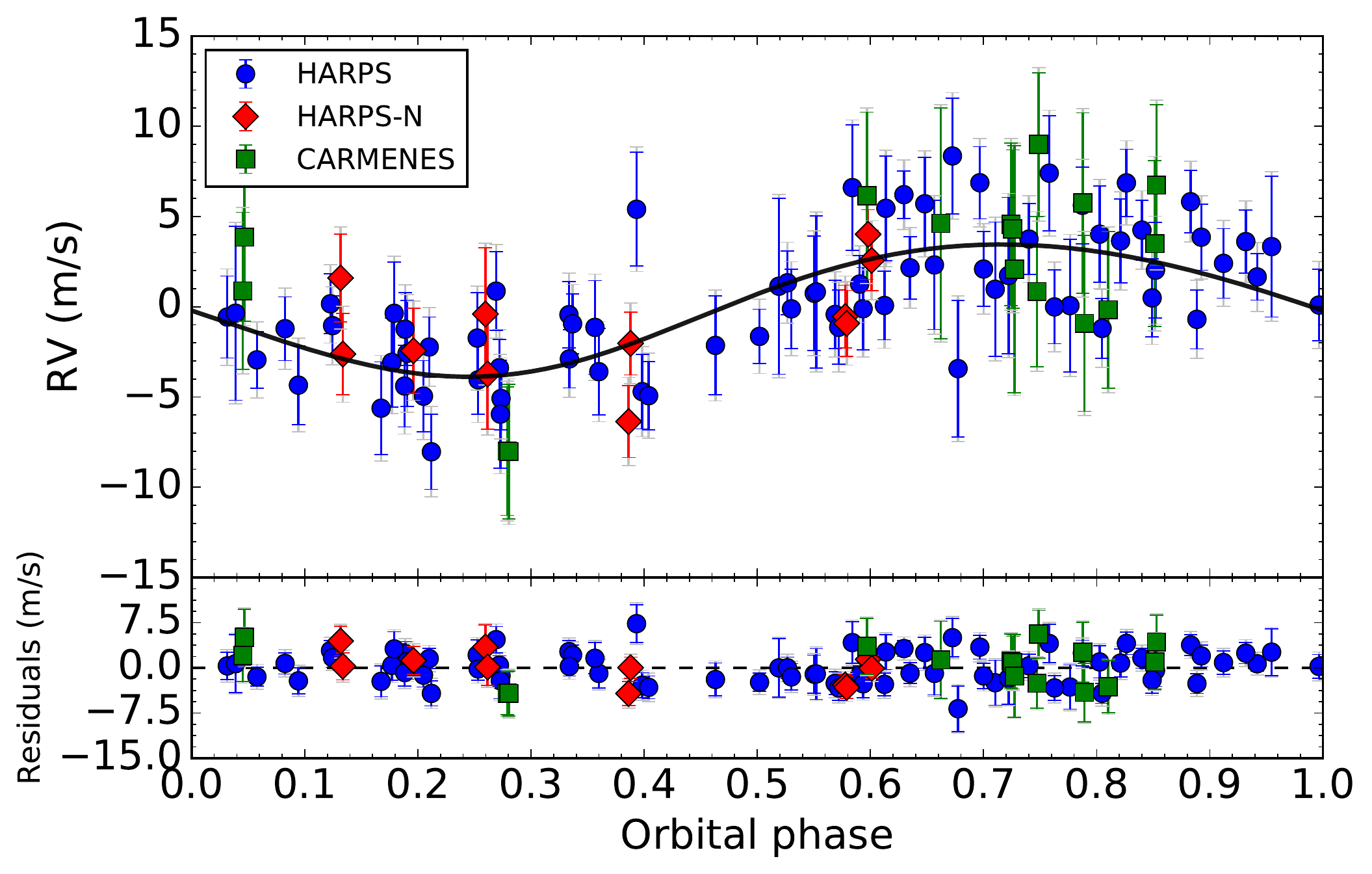}
  \includegraphics[width=9cm]{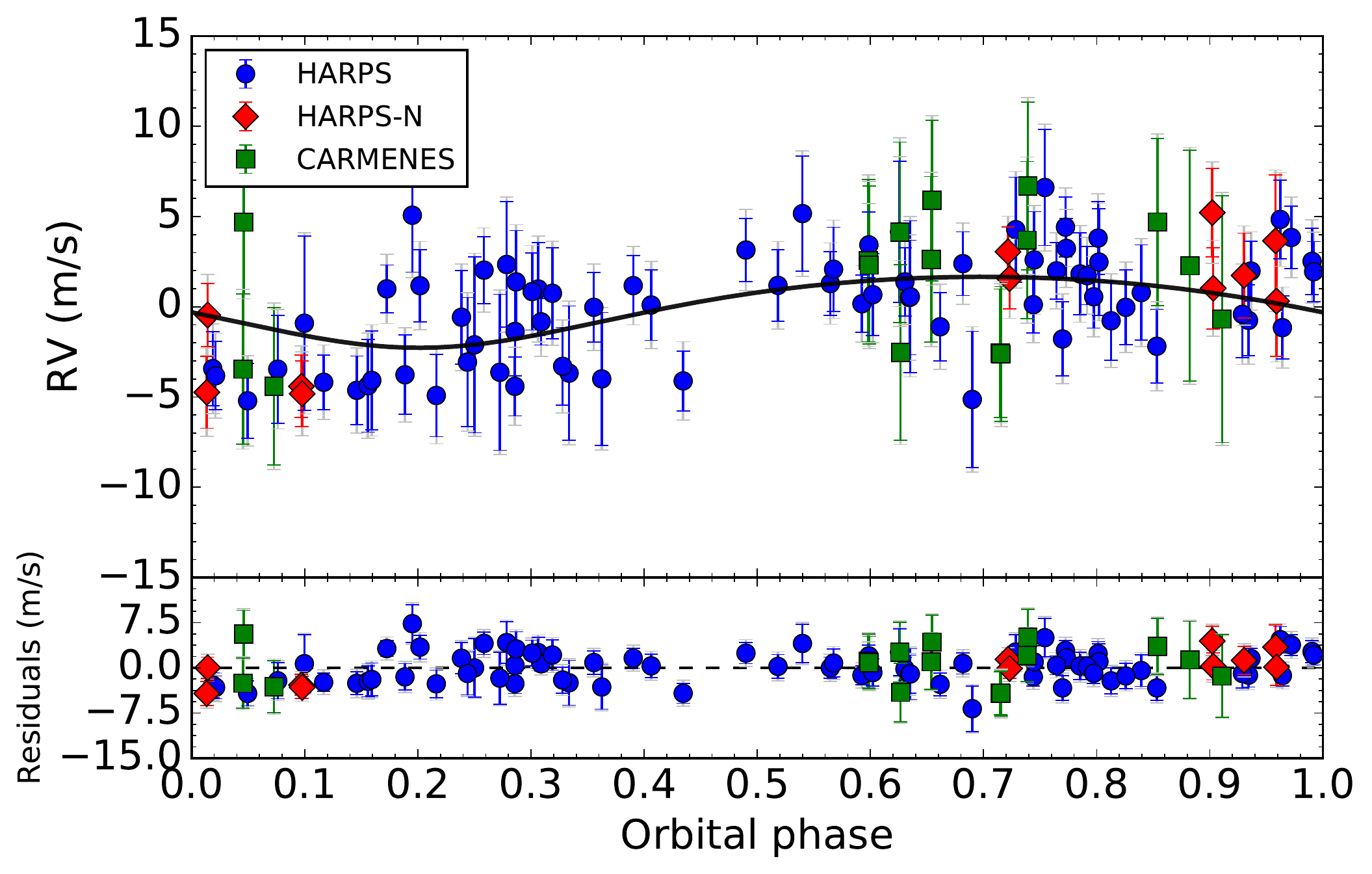}
  \includegraphics[width=9cm]{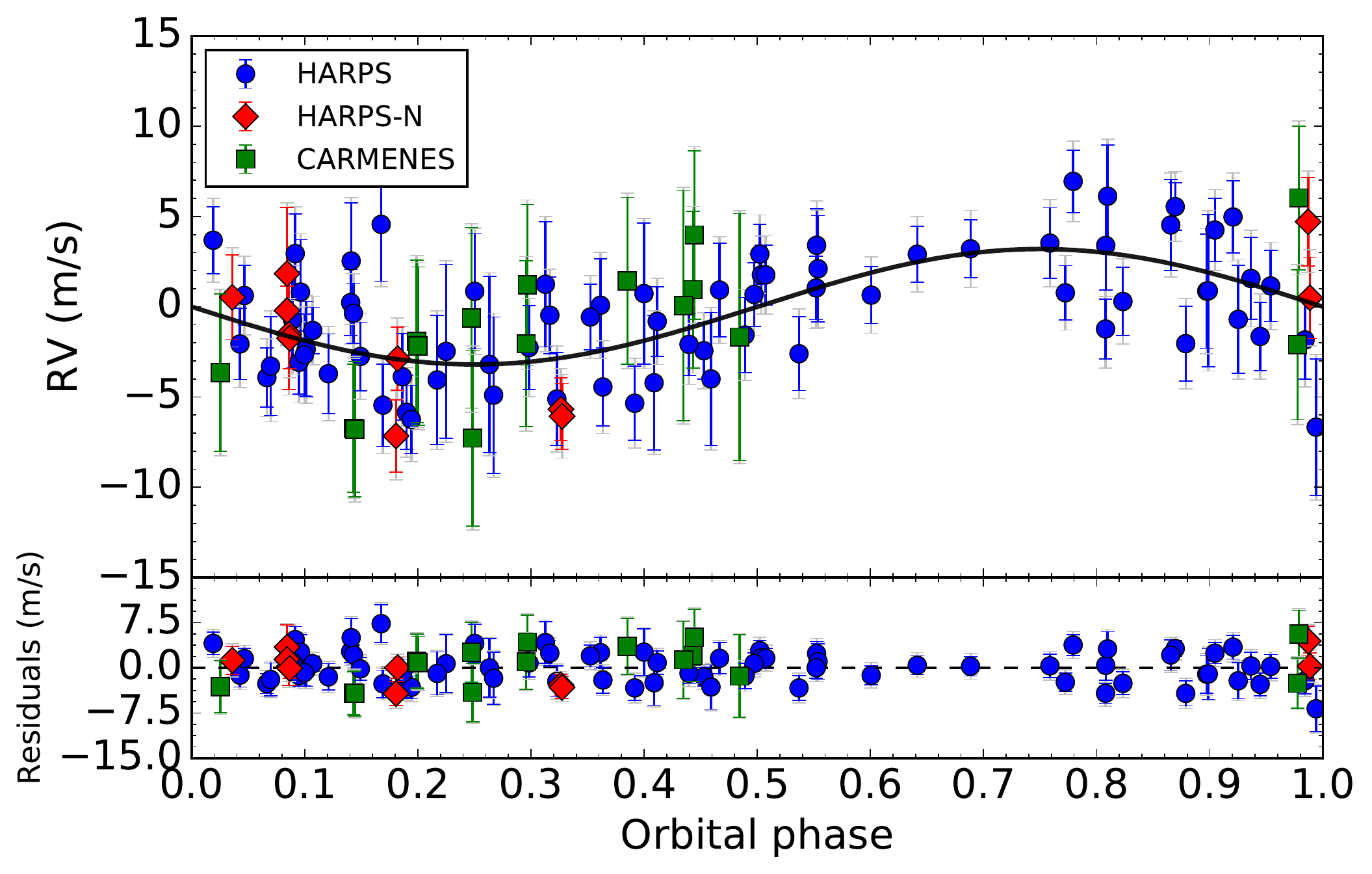}
     \label{fig:RVfinal}
     \caption{\textit{Top panel:} Time series of the RV measurements of \sname. Blue dots correspond to HARPS, red dots to HARPS-N, and green dots to CARMENES measurements. The best-fit model to the data is shown with a black thick line. The model includes three Keplerian curves and one sine curve mimicking the stellar signal at half the rotation period. \textit{Middle left panel:} Phase-folded RV measurements over the period of planet b after removing the signals from planets c and d and stellar activity. \textit{Middle right panel:} Phase-folded RV measurements over the period of planet c, after removing the signal from the other planets and stellar activity.  \textit{Bottom left panel:} Phase-folded RV measurements over the period of planet d, after subtracting the signal from planets b and c and stellar activity. \textit{Bottom right panel:} Phase-folded RV measurements over half the rotation period of the star after removing the signals from the three planets.}
\end{figure*}


\section{Joint analysis}
\label{sec:joint}

We simultaneously modeled the \ktwo\ transit photometry and HARPS, HARPS-N and CARMENES RV data with the software suite \texttt{pyaneti} \citep{Pyaneti}, which uses Markov chain Monte Carlo (MCMC) techniques to infer posterior distributions for the fitted parameters. The RV measurements were modeled using the sum of three Keplerian orbits and a sine signal at half the rotation period of the star (see Sect.~\ref{sec:freq}). The \ktwo\ transit light curves of the three planets were fit using the limb-darkened quadratic model of \citet{MandelAgol}. We integrated the light-curve model over ten steps to account for the 30\,min integration time of \ktwo\ \citep{2010MNRAS.408.1758K}. The fitted parameters are the systemic velocity $\gamma_{{\rm RV},i}$ for each instrument $i$, the RV semiamplitudes $K$, transit epochs $T_0$ and periods $P$ of the four Doppler signals, and the scaled semimajor axes $a/R_\star$, the planet-to-star radius ratios $R_p/R_\star$, the impact parameters $b$, the eccentricities $e$, the longitudes of periastron $\omega$, and the \citet{Kipping13} limb-darkening parametrization coefficients $q_1$ and $q_2$ for the three planets. We used the same expression for the likelihood as \citet{Barragan16} and created 500 independent chains for each parameter, using informative priors from our individual stellar, transit, and RV analyses to optimize computational time. Adequate convergence was considered when the Gelman–Rubin potential scale reduction factor dropped to within 1.03. After finding convergence, we ran 25\,000 more iterations with a thinning factor of 10, leading to a posterior distribution of 250\,000 independent samples for each fit parameter.

The orbital parameters and their uncertainties from our photometric and spectroscopic best joint fit model, are listed in Table~\ref{table:final}. They are defined as the median and 68\% region of the credible interval of the posterior distributions for each fit parameter. The resulting RV time series and phase-folded planetary signals are shown in Fig.~\ref{fig:RVfinal}. All three planets are detected at higher than the 3\,$\sigma$ level. The derived semiamplitudes for planets b, c, and d are 3.55$^{+0.43}_{-0.43}$\,m\,s$^{-1}$ , 3.66$^{+0.45}_{-0.46}$\,m\,s$^{-1}$, and 1.97$^{+0.54}_{-0.47}$\,m\,s$^{-1}$, respectively. The derived semiamplitude and period for the stellar activity signal are 3.20$^{+0.46}_{-0.47}$\,m\,s$^{-1}$ and 20.53$^{+0.04}_{-0.04}$ days.\newline

We also performed an independent joint analysis of our HARPS and HARPS-N RV and activity and symmetry indicator time series. We used the multidimensional Gaussian-process approach described by \citet{Rajpaul2015} as implemented in \texttt{pyaneti} by \citet[][]{Barragan2019b}. Briefly, this approach model RVs together with the activity and symmetry indicators assuming the same Gaussian process can describe them all following a quasi-periodic kernel. This approach has been used successfully to separate planet signals from stellar activity \citep[e.g.][]{Barragan2019b}. The inferred Doppler semiamplitudes for the three planets are consistent within 1\,$\sigma$ with the results presented in Table~\ref{table:final}. We also found the period of the quasi-periodic (QP) kernel to be P$_{\rm QP}\,=\,20.4\,\pm\,0.7$ d. This period comes from the correlation of the activity and symmetry indicators with the RV measurements, providing additional evidence that the $\sim$20-d RV signal is induced by stellar activity (see Sect.~\ref{sec:freq}).


\section{Discussion and conclusions}\label{sec:disc}

\begin{figure}
\centering
   \includegraphics[width=\linewidth]{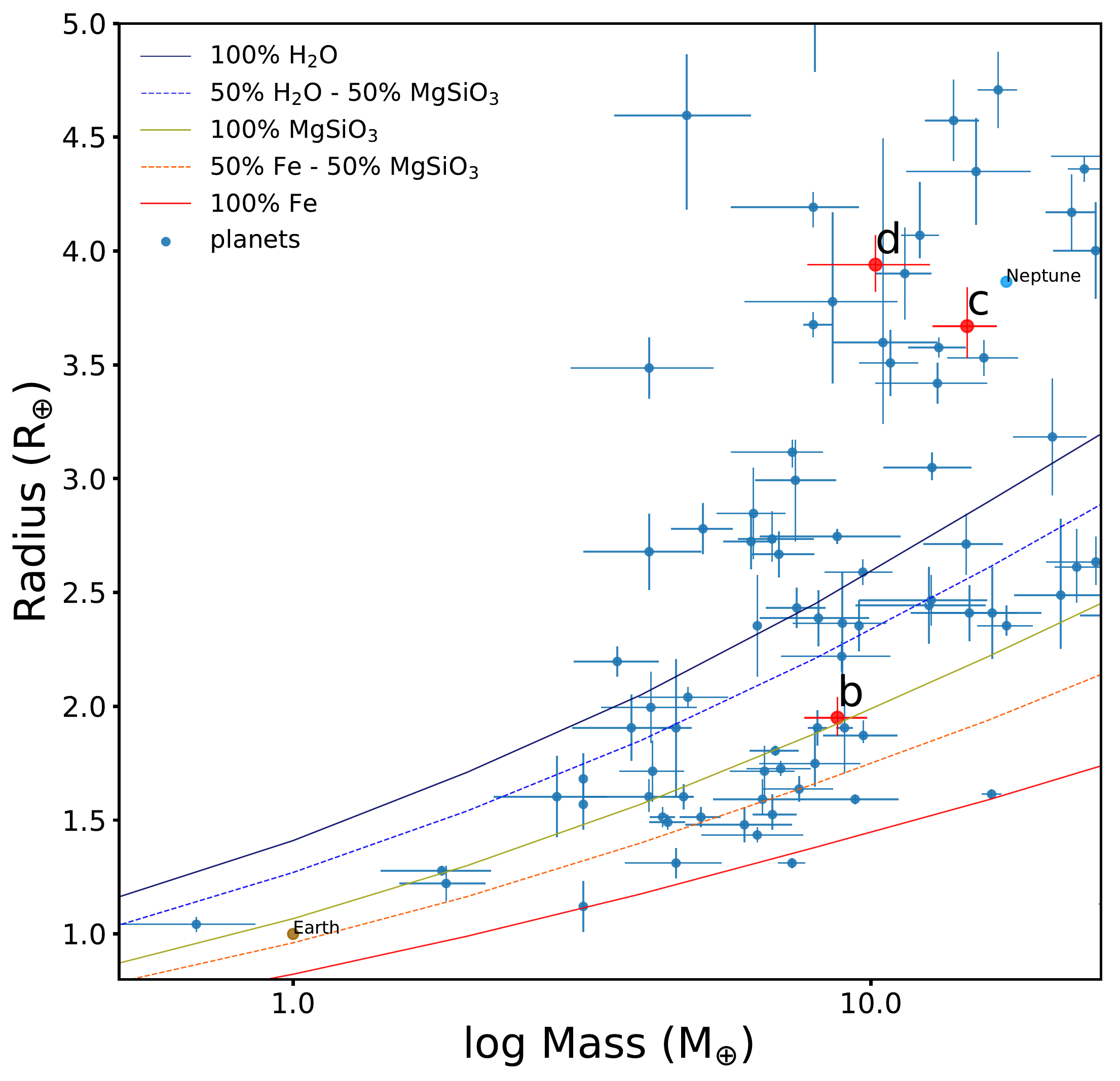}
     \caption{Mass vs. radius diagram for all known planets with masses in the range from 0.5 to 20.0~M$_{\oplus}$ and radii from 0.8 to 5.0~R$_{\oplus}$. Planets are shown only if the uncertainty in these two parameters is smaller than 30\%. Data are retrieved from the NASA Exoplanet Archive \protect\citep{2013PASP..125..989A} as of September 2019. Theoretical models for internal composition of small planets are taken from \protect\citep{Zeng2016}. The three planets we discovered and characterized in this paper are marked in red.
     }
     \label{fig:logM_R}
\end{figure}

We reported on the discovery of three small planets ($R_\mathrm{p}<4$\,$R_\odot$) transiting the evolved G8 IV/V star \sname. Combining \ktwo\ photometry with high-resolution imaging and high-precision Doppler spectroscopy, we confirmed the three planets and determined their masses, radii, and mean densities. With an orbital period of 3.6 days, the inner planet b, has a mass of $M_b\,=\,8.75^{+1.09}_{-1.08}\,\ M_{\oplus}$ and a radius of $R_b\,=\,1.95^{+0.09}_{-0.08}\,\ R_{\oplus}$, yielding a mean density of $\rho_b\,=\,6.39^{+1.19}_{-1.04}$~g\,cm$^{-3}$. With an orbital period of 15.6 days, planet c has a mass of $M_c\,=\,14.67^{+1.84}_{-1.89}\ M_{\oplus}$ and a radius of $R_c=3.67^{+0.17}_{-0.14}\ R_{\oplus}$, yielding a mean density of $\rho_c= 1.62^{+0.30}_{-0.29}$~g\,cm$^{-3}$. The outer planet d resides on a 35.7-day orbit, and has a mass of $M_d=10.18^{+2.46}_{-2.42}\,\ M_{\oplus}$ and a radius of $R_d\,=\,3.94^{+0.13}_{-0.12}~R_{\oplus}$, yielding a mean density of $\rho_d\,=\,0.91^{+0.25}_{-0.23}$~g\,cm$^{-3}$. For context, Figure~\ref{fig:logM_R} shows the mass-radius diagram for small planets with a mass and radius determination better than 30\%. The three new planets reported in this paper are also shown.

According to the \cite{Zeng2016} models, \sname\,b is a super-Earth with a density compatible with a pure silicate composition. However, a more realistic configuration would be a nickel-iron core and a silicate mantle. It lies above the model for 50\% iron - 50\% silicate, which probably means that it still has some residual H$_2$-He atmosphere, which enlarges its radius but does not significantly contribute to the total planet mass. As reported in \cite{Fulton2017} and \cite{VanEylen2018}, small planets follow a bimodal distribution with a valley at $\sim$1.5-2.2 $R_{\oplus}$ and peak at approximately 1.3 $R_{\oplus}$ for super-Earths and 2.4 $R_{\oplus}$ for sub-Neptunes. According to this, planet b, lies in the transition zone and might have lost most of its atmosphere through different mechanisms. The first is photoevaporation, which suggests a past atmosphere principally composed of hydrogen, which mainly occurs in the first 100 Myr of the stellar life, when it is more choromospherically active \citep{Owen2013}. On the other hand, \cite{Lee2014} proposed an alternative mechanism to explain a relatively thin atmosphere by delaying gas accretion into the planet until the gas in the protoplanetary disk is almost dissipated. Planetesimal impacts during planet formation can also encourage atmospheric loss \citep{Schlichting2015}, but it is unclear if impacts alone could produce the observed properties of planet b. \cite{Lopez2018} suggested that RV follow-up of long-period planets found in surveys such as \tess\ \citep{Ricker2015} or \plato\ \citep{Rauer2014} in the future should be able to distinguish between these two mechanisms because these two populations depend on semimajor-axis. Here, we estimate that given the proximity of planet b to its star ($\sim$0.05 AU), the influence of photoevaporation has been one of the most likely causes in the loss of its majority primordial hydrogen atmosphere.

\sname\ c and d are Neptune-sized planets, but with lower masses and hence lower mean densities (1.62 g cm$^{-3}$ and 0.91 g cm$^{-3}$ for planets c and d, respectively, vs. 1.95 g cm$^{-3}$ for Neptune), which suggest the presence of thicker atmospheres. Planet c has a stellar irradiation of $\sim$2.2$\cdot$10$^8$ erg cm$^{-2}$ s$^{-1}$, that is, slightly above the threshold of 2$\cdot$10$^8$ erg cm$^{-2}$ s$^{-1}$ established by \cite{Demory2011}, above which planets might inflate their atmospheres and be subject of photoevaporation. In contrast, planet d has a stellar irradiation of $\sim$7.2$\cdot$10$^7$ erg cm$^{-2}$ s$^{-1}$ and should in principle not be subjected to photoevaporation processes. The radius of planet c may therefore be compared to models in \cite{Fortney2007} for gas giant planets, based on which, we derive a core mass of $\sim$10 $M_{\oplus}$. The density, radius and mass of planet d suggest a relatively small but heavy core with a thick atmosphere.

Based on the study of three planetary systems, \citet{Grunblatt2018} proposed that close-in planets orbiting evolved stars tend to reside on eccentric orbits. If this scenario is correct, the nearly circular orbits of \sname\,b, c, and d may be the result of the planets orbiting a star that is not evolved enough for a fair comparison to be made. According to the distance deduced in Table~\ref{table:final}, we consider the three planets of the system EPIC 249893012 as close-in planets, with circular orbits, although for planet d a wide range of eccentricities from 0.04 to 0.36 is possible. Because the system is at an early stage of its evolution after leaving the main sequence, it is a good candidate for a detailed study of its dynamical evolution, to i) shed light on  the formation of close-in giant planets \citep{Dawson2018}, and ii) test the hypothesis by \cite{Izidoro2015} that giant planets form a dynamical barrier that confines super-Earths to an inward-migrating evolution.

\begin{acknowledgements}
D.H. acknowledges the Spanish Ministry of Economy and Competitiveness (MINECO) for the financial support under the FPI programme BES-2015-075200. This work is partly financed by the Spanish Ministry of Economics and Competitiveness through project ESP2016-80435-C2-2-R.  This work is partly supported by JSPS KAKENHI Grant Numbers JP18H01265 and JP18H05439, and JST PRESTO Grant Number JPMJPR1775. This work was supported by the KESPRINT collaboration, an international consortium devoted to the characterization and research of exoplanets discovered with space-based missions. SM acknowledges support from the Spanish Ministry under the Ramon y Cajal fellowship number RYC-2015-17697. HJD and GN acknowledge support by grant ESP2017-87676-C5-4-R of the Spanish Secretary of State for R\&D\&i (MINECO). SzCs, ME, SG, APH, JK, KWFL, MP and HR acknowledge support by DFG grants PA525/18-1, PA525/19-1, PA525/20-1, HA3279/12-1 and RA714/14-1 within the DFG Schwerpunkt SPP 1992, "Exploring the Diversity of Extrasolar Planets". ICL acknowledges CNPq, CAPES, and FAPERN Brazilian agencies. This research has made use of the NASA Exoplanet Archive, which is operated by the California Institute of Technology, under contract with the National Aeronautics and Space Administration under the Exoplanet Exploration Program. This work has made use of data from the European Space Agency (ESA) mission {\it Gaia}\footnote{\url{https://www.cosmos.esa.int/gaia}}, processed by the {\it Gaia} Data Processing and Analysis Consortium (DPAC\footnote{\url{https://www.cosmos.esa.int/web/gaia/dpac/consortium}}). Funding for the DPAC has been provided by national institutions, in particular the institutions participating in the {\it Gaia} Multilateral Agreement.

This paper has made use of data collected by the \ktwo\ mission. Funding for the \ktwo\ mission is provided by the NASA Science Mission directorate.
\end{acknowledgements}

\bibliographystyle{aa}
\balance
\bibliography{biblio}

\begin{table*}[t!]
  {\renewcommand{\arraystretch}{1.25}
\label{table:final}
\centering
\caption{Parameters of the three planets and stellar signal from the joint-analysis fit.}
\begin{tabular}{l|ccccc}
\hline\hline
\noalign{\smallskip}
Parameter & Planet b & Planet c & Planet d & Stellar signal \\
\noalign{\smallskip}
\textbf{Transit and RV model parameters} & & & &   \\
Orbital period $P_{\textrm{orb}}\ (d)$              & 3.5951$_{-0.0003}^{+0.0003}$  & 15.624$_{-0.001}^{+0.001}$    & 35.747$_{-0.005}^{+0.005}$    & 20.53$_{-0.04}^{+0.04}$   \\
Epoch $T_0$ (BJD$_\mathrm{TDB}$ $-$ 2454833; d)                        & 3161.396$_{-0.005}^{+0.005}$  & 3165.841$_{-0.004}^{+0.002}$  & 3175.652$_{-0.003}^{+0.003}$  & 3263.72$_{-0.95}^{+0.86}$   \\
Scaled semimajor axis $a/R_*$                          & 5.93$_{-0.60}^{+0.96}$        & 15.79$_{-2.56}^{+1.58}$       & 27.42$_{-4.44}^{+2.74}$       & \dots   \\
Planet-to-star ratio radius $r_p/R_*$                      & 0.0104$_{-0.0004}^{+0.0004}$  & 0.0197$_{-0.0006}^{+0.0008}$  & 0.0211$_{-0.0004}^{+0.0005}$  & \dots   \\
Impact parameter $b$                                & 0.42$_{-0.25}^{+0.28}$        & 0.60$_{-0.21}^{+0.15}$        & 0.25$_{-0.17}^{+0.23}$        & \dots   \\
$\sqrt{e}\sin \omega_*$                             &-0.08$_{-0.23}^{+0.24}$        &-0.02$_{-0.26}^{+0.25}$        &-0.01$_{-0.27}^{+0.29}$        & 0   \\
$\sqrt{e}\cos \omega_*$                             &-0.04$_{-0.16}^{+0.16}$        &0.12$_{-0.18}^{+0.12}$         &-0.23$_{-0.30}^{+0.29}$        & 0   \\
Doppler semiamplitude $K\ (m\ s^{-1})$             & 3.55$_{-0.43}^{+0.43}$        & 3.66$_{-0.46}^{+0.45}$        & 1.97$_{-0.47}^{+0.54}$        & 3.20$_{-0.47}^{+0.46}$   \\
Parameterized limb-darkening coefficient $q_1$  & $0.43\pm0.09$ &\dots &\dots &\dots \\
Parameterized limb-darkening coefficient $q_2$  & $0.22\pm0.09$ &\dots &\dots &\dots \\
Systemic velocity $\gamma_\mathrm{HARPS}\ (km\ s^{-1})$    & 21.6127$_{-0.0003}^{+0.0003}$ &\dots&\dots&\dots     \\
Systemic velocity $\gamma_\mathrm{HARPS-N}\ (km\ s^{-1})$  & 21.6080$_{-0.0009}^{+0.0009}$ &\dots&\dots&\dots     \\
Systemic velocity $\gamma_\mathrm{CARMENES}\ (km\ s^{-1})$ & 49.660$_{-0.001}^{+0.001}$ &\dots&\dots&\dots     \\
RV jitter $\sigma_\mathrm{HARPS}\ (m\ s^{-1})$             & 1.40$_{-0.42}^{+0.43}$ &\dots&\dots&\dots     \\
RV jitter $\sigma_\mathrm{HARPS-N}\ (m\ s^{-1})$           & 1.41$_{-1.29}^{+0.95}$ &\dots&\dots&\dots     \\
RV jitter $\sigma_\mathrm{CARMENES}\ (m\ s^{-1})$          & 1.51$_{-1.53}^{+1.05}$ &\dots&\dots&\dots     \\
\noalign{\smallskip}
\hline
\noalign{\smallskip}
\textbf{Derived parameters} & & & &    \\
Planet radius $R_\mathrm{p}\ (R_{\oplus})$                       & 1.95$_{-0.08}^{+0.09}$    & 3.67$_{-0.14}^{+0.17}$    & 3.94$_{-0.12}^{+0.13}$ & \dots   \\
Planet mass $M_\mathrm{p}\ (M_{\oplus})$                         & 8.75$_{-1.08}^{+1.09}$    & 14.67$_{-1.89}^{+1.84}$   & 10.18$_{-2.42}^{+2.46}$ & \dots   \\
Planet density $\rho_\mathrm{p}\ (g\ cm^{-3})$                   & 6.39$_{-1.04}^{+1.19}$    & 1.62$_{-0.29}^{+0.30}$    & 0.91$_{-0.23}^{+0.25}$ & \dots   \\
Time of periastron passage ($d$)                        &3161.67$_{-1.7}^{+1.2}$    & 3165.3$_{-3.7}^{+4.4}$ & 3175.77$_{-9.0}^{+7.9}$ & \dots   \\
Semimajor axis $a\ (AU)$                        &0.047$_{-0.007}^{+0.005}$  & 0.13$_{-0.02}^{+0.01}$    & 0.22$_{-0.04}^{+0.02}$ & \dots   \\
Orbit inclination  $i_\mathrm{p}$ $deg$                     & 86.14$_{-3.50}^{+2.60}$   & 87.94$_{-1.05}^{+0.74}$   & 89.47$_{-0.50}^{+0.36}$ & \dots   \\
Eccentricity $e$                                        &0.06$_{-0.04}^{+0.08}$     &0.07$_{-0.05}^{+0.08}$     &0.15$_{-0.11}^{+0.21}$ & \dots   \\
Longitude of periastron $\omega_*\ (^{\circ})$          &225$_{-123}^{+67}$         &217$_{-170}^{+100}$        &181$_{-61}^{+81}$ & \dots   \\
Transit duration $\tau_{14}\ (h)$                       &4.33$_{-0.15}^{+0.18}$     & 6.37$_{-0.12}^{+0.15}$    & 9.56$_{-0.13}^{+0.14}$ & \dots   \\
Equilibrium temperature $T_{\textrm{eq}}\ (\textrm{K})$ & 1616$_{-79}^{+149}$       & 990$_{-49}^{+92}$         & 752$_{-37}^{+69}$ & \dots   \\
Insolation $F\ (F_{\oplus})$                            & 1037$_{-207}^{+482}$      & 160$_{-29}^{+68}$         & 53$_{-10}^{+23}$ & \dots   \\
\hline
\end{tabular}}
\end{table*}

\clearpage

\begin{table*}
\caption{HARPS  measurements of \sname.}
\label{table:RV-H}
\centering
\scriptsize
\begin{tabular}{l|ccrcrcccc}
\hline
\hline
\noalign{\smallskip}
BJD$_{TDB}$ & RV & eRV & CCF$_\mathrm{BIS}$ & CCF$_\mathrm{FWHM}$ & log R$_\mathrm{HK}$ & elog R$_\mathrm{HK}$ & S/N$_{5500\,\AA}$ & T$_\mathrm{exp}$ & Instrument \\
(d) & (km s$^{-1}$) & (km s$^{-1}$) & (km s$^{-1}$)& (km s$^{-1}$)& & & & (s) &  \\
\noalign{\smallskip}
\hline
\noalign{\smallskip}
2458220.78441&21.6078&0.0020&-0.0075&7.1754&-5.18&0.04&40.8&2400&HARPS\\
2458220.86371&21.6099&0.0017&-0.0089&7.1837&-5.19&0.03&48.9&2400&HARPS\\
2458221.78704&21.6083&0.0022& 0.0004&7.1679&-5.16&0.04&37.2&2400&HARPS\\
2458221.85556&21.6023&0.0018&-0.0132&7.1876&-5.17&0.03&45.9&2400&HARPS\\
2458222.79324&21.6108&0.0018&-0.0009&7.1885&-5.17&0.03&44.1&2400&HARPS\\
2458222.84368&21.6106&0.0017&-0.0152&7.1921&-5.20&0.03&48.3&2400&HARPS\\
2458223.80514&21.6074&0.0021&-0.0058&7.1831&-5.17&0.04&40.0&2400&HARPS\\
2458223.89387&21.6068&0.0019&-0.0016&7.1926&-5.17&0.04&43.9&2400&HARPS\\
2458249.73305&21.6098&0.0016&-0.0051&7.1866&-5.18&0.03&50.3&2700&HARPS\\
2458250.74892&21.6110&0.0017&-0.0069&7.1858&-5.22&0.04&48.2&2400&HARPS\\
2458250.77753&21.6099&0.0017&-0.0047&7.1945&-5.15&0.03&48.7&2400&HARPS\\
2458251.77954&21.6163&0.0020&-0.0086&7.1831&-5.14&0.04&40.7&2400&HARPS\\
2458251.80647&21.6151&0.0029& 0.0011&7.1821&-5.05&0.06&29.6&2400&HARPS\\
2458324.63972&21.6164&0.0026&-0.0053&7.1896&-5.25&0.07&32.4&2400&HARPS\\
2458543.86587&21.6243&0.0017&-0.0034&7.1865&-5.36&0.07&51.3&2400&HARPS\\
2458551.83867&21.6117&0.0032&-0.0072&7.1884&-5.23&0.09&30.6&2400&HARPS\\
2458553.80415&21.6111&0.0049&-0.0010&7.1938&-5.78&0.59&22.0&2400&HARPS\\
2458554.81840&21.6165&0.0035& 0.0105&7.1750&-5.30&0.14&28.9&2400&HARPS\\
2458555.81881&21.6113&0.0026&-0.0087&7.1888&-5.58&0.17&36.6&2400&HARPS\\
2458556.79144&21.6097&0.0037& 0.0036&7.1912&-5.20&0.11&27.0&2400&HARPS\\
2458557.82395&21.6145&0.0037&-0.0128&7.1781&-5.50&0.21&27.2&2400&HARPS\\
2458558.81788&21.6165&0.0017&-0.0082&7.1929&-5.27&0.05&53.2&2400&HARPS\\
2458570.90587&21.6153&0.0029&-0.0076&7.1910&-5.11&0.07&33.8&2700&HARPS\\
2458571.82537&21.6229&0.0032&-0.0043&7.1805&-5.15&0.08&30.8&2700&HARPS\\
2458584.78548&21.6088&0.0015&-0.0075&7.1924&-5.17&0.03&56.9&2400&HARPS\\
2458585.83025&21.6139&0.0019&-0.0041&7.1889&-5.16&0.04&45.9&2700&HARPS\\
2458586.77962&21.6222&0.0013&-0.0086&7.1878&-5.14&0.02&64.4&2700&HARPS\\
2458587.82822&21.6173&0.0020& 0.0017&7.1815&-5.12&0.04&44.6&2400&HARPS\\
2458589.85191&21.6200&0.0019&-0.0116&7.1851&-5.12&0.03&46.2&2700&HARPS\\
2458590.82593&21.6106&0.0016&-0.0113&7.2020&-5.13&0.03&52.6&2700&HARPS\\
2458591.65814&21.6081&0.0013&-0.0178&7.2019&-5.17&0.02&64.3&2700&HARPS\\
2458600.79471&21.6196&0.0018&-0.0050&7.1874&-5.28&0.06&52.7&2700&HARPS\\
2458601.79254&21.6181&0.0016&-0.0014&7.2017&-5.29&0.05&57.7&2700&HARPS\\
2458608.68407&21.6171&0.0023&-0.0004&7.1847&-5.31&0.07&40.6&2700&HARPS\\
2458609.66596&21.6072&0.0022&-0.0021&7.1933&-5.16&0.05&42.9&2700&HARPS\\
2458613.81491&21.6031&0.0024&-0.0049&7.2012&-5.24&0.09&40.2&2700&HARPS\\
2458626.58042&21.6134&0.0025&-0.0071&7.1917&-5.11&0.04&34.7&2700&HARPS\\
2458626.61341&21.6159&0.0028&-0.0023&7.1932&-5.16&0.06&31.0&2700&HARPS\\
2458627.75972&21.6087&0.0025&-0.0102&7.1941&-5.37&0.12&39.0&2700&HARPS\\
2458635.65975&21.6156&0.0032&-0.0170&7.2012&-5.21&0.09&31.0&2700&HARPS\\
2458636.64443&21.6176&0.0023&-0.0094&7.2018&-5.26&0.07&39.8&2700&HARPS\\
2458637.75899&21.6162&0.0018&-0.0058&7.1791&-5.34&0.07&51.0&2700&HARPS\\
2458638.73082&21.6120&0.0039& 0.0075&7.1803&-5.16&0.13&27.6&2700&HARPS\\
2458640.73000&21.6162&0.0018&-0.0004&7.1927&-5.19&0.05&52.3&2700&HARPS\\
2458643.69045&21.6159&0.0016&-0.0064&7.1887&-5.42&0.08&58.5&2400&HARPS\\
2458644.65679&21.6164&0.0016&-0.0049&7.1905&-5.19&0.05&56.7&2400&HARPS\\
2458655.67055&21.6086&0.0048& 0.0051&7.1902&-4.87&0.06&21.0&2400&HARPS\\
2458657.68053&21.6025&0.0026&-0.0061&7.1712&-5.12&0.05&35.0&2400&HARPS\\
2458660.63404&21.6054&0.0026&-0.0008&7.1882&-5.08&0.05&34.5&2400&HARPS\\
2458666.62842&21.6195&0.0020&-0.0119&7.1968&-5.12&0.04&44.5&2400&HARPS\\
2458667.63837&21.6109&0.0017&-0.0065&7.1932&-5.13&0.03&52.4&2700&HARPS\\
2458669.62253&21.6217&0.0018&-0.0032&7.1965&-5.17&0.03&49.4&2400&HARPS\\
2458670.63466&21.6125&0.0020&-0.0093&7.1995&-5.13&0.04&44.1&2400&HARPS\\
2458673.63051&21.6090&0.0023&-0.0076&7.1913&-5.14&0.04&38.4&2680&HARPS\\
2458674.64597&21.6040&0.0019&-0.0087&7.1856&-5.17&0.04&46.6&2400&HARPS\\
2458679.62480&21.6136&0.0021&-0.0091&7.2000&-5.56&0.17&47.3&2400&HARPS\\
2458680.61023&21.6173&0.0017&-0.0058&7.1990&-5.34&0.07&54.3&2400&HARPS\\
2458681.62554&21.6128&0.0021&-0.0133&7.1888&-5.27&0.07&46.2&2400&HARPS\\
2458682.60541&21.6134&0.0020&-0.0083&7.1965&-5.27&0.06&45.6&2400&HARPS\\
2458689.62388&21.6037&0.0021&-0.0065&7.1943&-5.33&0.10&46.1&2400&HARPS\\
2458690.57672&21.6099&0.0030&-0.0036&7.1948&-5.31&0.11&32.8&2400&HARPS\\
2458693.54664&21.6069&0.0027&-0.0129&7.1939&-5.22&0.09&36.0&2400&HARPS\\
2458694.59633&21.6114&0.0022&-0.0027&7.1935&-5.29&0.07&43.0&2400&HARPS\\
2458695.58627&21.6065&0.0023&-0.0130&7.1918&-5.54&0.13&41.5&2400&HARPS\\
2458696.57103&21.6064&0.0036& 0.0020&7.2044&-5.44&0.21&29.7&2400&HARPS\\
2458697.59484&21.6118&0.0043&-0.0039&7.1721&-5.84&0.74&25.6&2400&HARPS\\
2458698.61455&21.6158&0.0021&-0.0093&7.1924&-5.24&0.07&44.7&2400&HARPS\\
2458699.57794&21.6065&0.0022&-0.0125&7.1905&-5.33&0.09&44.2&2400&HARPS\\
2458700.56352&21.6112&0.0019&-0.0093&7.1829&-5.41&0.08&48.1&2400&HARPS\\
2458711.50542&21.6153&0.0019& 0.0008&7.1904&-5.36&0.08&49.6&2400&HARPS\\
2458712.52173&21.6144&0.0038&-0.0275&7.1965&-5.55&0.27&28.3&2400&HARPS\\
2458714.47713&21.6131&0.0027&-0.0094&7.1841&-5.14&0.07&36.5&2100&HARPS\\
\noalign{\smallskip}
\hline
\end{tabular}
\end{table*}

\begin{table*}[t]
\caption{HARPS-N measurements of \sname.}
\label{table:RV-HN}
\centering
\scriptsize
\begin{tabular}{l|ccrcrcccc}
\hline
\hline
\noalign{\smallskip}
BJD$_\mathrm{TDB}$ & RV & eRV & CCF$_\mathrm{BIS}$ & CCF$_\mathrm{FWHM}$ & log R$_\mathrm{HK}$ & elog R$_\mathrm{HK}$ & S/N$_{5500\,\AA}$ & T$_\mathrm{exp}$ & Instrument \\
(d) & (km s$^{-1}$) & (km s$^{-1}$) & (km s$^{-1}$)& (km s$^{-1}$)& & & & (s) &  \\
\noalign{\smallskip}
\hline
\noalign{\smallskip}
2458219.63943&21.6138&0.0024&-0.0087&7.1193&-5.11&0.05&38.1&2700&HARPS-N\\
2458219.67084&21.6096&0.0023&-0.0173&7.1197&-5.15&0.05&40.6&2700&HARPS-N\\
2458220.64565&21.6066&0.0024&-0.0083&7.1229&-5.11&0.05&39.1&2700&HARPS-N\\
2458221.63935&21.6026&0.0037&-0.0157&7.1333&-5.14&0.09&27.9&2600&HARPS-N\\
2458221.66954&21.5992&0.0031&-0.0114&7.1269&-5.18&0.08&31.8&2600&HARPS-N\\
2458223.61735&21.6016&0.0020&-0.0069&7.1224&-5.19&0.05&45.2&2400&HARPS-N\\
2458223.64464&21.6059&0.0017&-0.0069&7.1208&-5.16&0.04&50.2&2400&HARPS-N\\
2458226.60910&21.6055&0.0017&-0.0134&7.1246&-5.17&0.03&50.3&1800&HARPS-N\\
2458226.62995&21.6052&0.0018&-0.0163&7.1258&-5.19&0.04&48.2&1800&HARPS-N\\
2458570.66046&21.6085&0.0014&-0.0149&7.1242&-5.16&0.02&62.4&3600&HARPS-N\\
2458570.70977&21.6072&0.0017&-0.0120&7.1222&-5.20&0.04&51.7&3600&HARPS-N\\
\noalign{\smallskip}
\hline
\end{tabular}
\end{table*}

\begin{table*}
\caption{CARMENES  measurements of \sname.}
\label{table:RV-C}
\centering
\scriptsize
\begin{tabular}{l|ccrcccccc}
\hline
\hline
\noalign{\smallskip}
BJD$_\mathrm{TDB}$ & RV & eRV & CCF$_\mathrm{BIS}$ & CCF$_\mathrm{FWHM}$ & log R$_\mathrm{HK}$ & elog R$_\mathrm{HK}$ & S/N$_{5340\,\AA}$ & T$_\mathrm{exp}$ & Instrument \\
(d) & (km s$^{-1}$) & (km s$^{-1}$) & (km s$^{-1}$)& (km s$^{-1}$)& & & & (s) &  \\
\noalign{\smallskip}
\hline
\noalign{\smallskip}
2458244.52311&49.6652&0.0045&-0.0132&7.7551&--&--&59.9&1800&CARMENES\\
2458244.54617&49.6651&0.0044&-0.0079&7.7543&--&--&59.6&1800&CARMENES\\
2458245.51467&49.6667&0.0050& 0.0026&7.7540&--&--&53.6&1800&CARMENES\\
2458245.53632&49.6600&0.0049&-0.0253&7.7772&--&--&56.5&1800&CARMENES\\
2458246.50860&49.6596&0.0046&-0.0458&7.7732&--&--&58.8&1800&CARMENES\\
2458246.53124&49.6627&0.0045&-0.0338&7.7594&--&--&60.8&1800&CARMENES\\
2458249.53854&49.6620&0.0044&-0.0081&7.7522&--&--&62.1&1800&CARMENES\\
2458249.56019&49.6649&0.0047&-0.0248&7.7298&--&--&58.0&1800&CARMENES\\
2458260.50626&49.6600&0.0042&-0.0310&7.7646&--&--&64.2&1800&CARMENES\\
2458260.52919&49.6680&0.0040&-0.0256&7.7473&--&--&67.5&1800&CARMENES\\
2458261.48996&49.6547&0.0044&-0.0183&7.7661&--&--&60.3&1800&CARMENES\\
2458284.43860&49.6546&0.0036&-0.0222&7.7784&--&--&75.8&1800&CARMENES\\
2458284.46094&49.6546&0.0037&-0.0389&7.7531&--&--&71.8&1800&CARMENES\\
2458289.40632&49.6640&0.0046&-0.0244&7.7711&--&--&59.8&1800&CARMENES\\
2458290.42667&49.6615&0.0064& 0.0080&7.7874&--&--&44.2&1800&CARMENES\\
2458291.44303&49.6655&0.0068&-0.0469&7.7647&--&--&41.2&1800&CARMENES\\
\noalign{\smallskip}
\hline
\end{tabular}
\end{table*}

\end{document}